\documentclass[a4paper, journal]{IEEEtran}
\pdfoutput=1

\usepackage[T1]{fontenc}
\usepackage[utf8]{inputenc}

\usepackage{amsmath}
\usepackage[cmintegrals]{newtxmath}
\usepackage{bm}
\usepackage{cite}
\usepackage{color}
\usepackage{graphicx}
\usepackage{flushend}
\usepackage[acronym,toc,shortcuts]{glossaries-extra}
\usepackage{booktabs}
\usepackage{dcolumn}
\usepackage{orcidlink}
\usepackage{xurl}
\usepackage{hyperref}

\usepackage{multirow}
\usepackage{pifont}

\hypersetup{
	pdfauthor={Maximilian Engelhardt, Sebastian Giehl, Michael Schubert, Alexander Ihlow, Christian Schneider, Alexander Ebert, Markus Landmann, Giovanni Del Galdo, Carsten Andrich},
	pdftitle={Accelerating Innovation in 6G Research: Real-Time Capable SDR System Architecture for Rapid Prototyping},
    pdfkeywords={6G, integrated sensing and communication (ISAC), rapid prototyping, RFSoC, software-defined radio (SDR), system architecture, wideband streaming and processing}
}

\usepackage{siunitx}
\usepackage{soul}
\usepackage{subcaption}
\usepackage{tablefootnote}
\usepackage{makecell}
\DeclareSIUnit \belc {Bc}
\DeclareSIUnit{\belmilliwatt}{Bm}
\DeclareSIUnit{\dBm}{\deci\belmilliwatt}
\DeclareSIUnit{\dBc}{\deci\belc}
\DeclareSIUnit{\sample}{\text{Sa}}

\usepackage{placeins}

\usepackage{tikz}
\usetikzlibrary{positioning}
\usetikzlibrary{shapes}
\usetikzlibrary{decorations.markings, spy, fit, patterns, arrows}
\usetikzlibrary{intersections}
\usepackage{circuitikz}
\tikzset{>=latex}
\pgfdeclarelayer{bg}
\pgfdeclarelayer{fg}
\pgfsetlayers{bg,main,fg}

\usepackage{pgfplots}
\usepgfplotslibrary{fillbetween}
\pgfplotsset{compat=1.16}
\usepackage{glossaries}

\usepackage{relsize}
\usepackage{xspace}
\newcommand{\Rplus}{\protect\hspace{-.1em}\protect\raisebox{.35ex}{\smaller{\smaller\textbf{+}}}}
\newcommand{\Cpp}{\mbox{C\Rplus\Rplus}\xspace}

\setabbreviationstyle[acronym]{long-short}
\makeglossaries

\newacronym{FPGA}{FPGA}{field-programmable gate array}
\newacronym{PC}{PC}{personal computer}
\newacronym{SDR}{SDR}{software-defined radio}
\newacronym{Rx}{Rx}{receive}
\newacronym{Tx}{Tx}{transmit}
\newacronym{ADC}{ADC}{analog-to-digital converter}
\newacronym{DAC}{DAC}{digital-to-analog converter}
\newacronym{PEP}{PEP}{peak envelope power}
\newacronym{USRP}{USRP}{universal software radio peripheral}
\newacronym{DUT}{DUT}{device under test}
\newacronym{SNR}{SNR}{signal-to-noise ratio}
\newacronym{MIMO}{MIMO}{multiple input multiple output}
\newacronym{DPDK}{DPDK}{Data Plane Development Kit}
\newacronym{NIC}{NIC}{network interface controller}
\newacronym{GPU}{GPU}{graphics processing unit}
\newacronym{CPU}{CPU}{central processing unit}
\newacronym{AXI}{AXI4}{AMBA advanced eXtensible interface 4}
\newacronym{AXIoE}{AXIoE}{AXI over Ethernet}
\newacronym{LO}{LO}{local oscillator}
\newacronym{RF}{RF}{radio frequency}
\newacronym{COTS}{COTS}{commercial off-the-shelf}
\newacronym{RFdc}{RFdc}{RF data converter}
\newacronym{DDC}{DDC}{digital down conversion}
\newacronym{DUC}{DUC}{digital up conversion}
\newacronym{AGC}{AGC}{automatic gain control}
\newacronym{FIFO}{FIFO}{first in, first out}
\newacronym{PS}{PS}{processing system}
\newacronym{PL}{PL}{user-programmable logic}
\newacronym{MTS}{MTS}{multi-tile synchronization}
\newacronym{CHDR}{CHDR}{condensed hierarchical datagram for RFNoC}
\newacronym{ISAC}{ISAC}{integrated sensing and communication}
\newacronym{HRR}{HRR}{high range resolution}
\newacronym{VLAN}{VLAN}{virtual local area network}
\newacronym{SPI}{SPI}{serial peripheral interface}
\newacronym{I2C}{I\textsuperscript{2}C}{inter-integrated circuit}
\newacronym{CRC}{CRC}{cyclic redundancy check}
\newacronym{RS-FEC}{RS-FEC}{Reed-Solomon forward error correction}
\newacronym{RPC}{RPC}{remote procedure call}
\newacronym{TCP}{TCP}{transmission control protocol}
\newacronym{UDP}{UDP}{user datagram protocol}
\newacronym{RAID}{RAID}{redundant array of independent disks}
\newacronym{MTU}{MTU}{maximum transmission unit}
\newacronym{SYSREF}{SYSREF}{system synchronization reference signal}
\newacronym{REFCLK}{REFCLK}{reference clock}
\newacronym{RTC}{RTC}{real time clock}
\newacronym{sota}{SotA}{state-of-the-art}
\newacronym{UHD}{UHD}{USRP Hardware Driver}
\newacronym{BRAM}{BRAM}{block random access memory}
\newacronym{RoCE}{RoCE}{RDMA over converged Ethernet}
\newacronym{RFNoC}{RFNoC}{RF Network on Chip}

\pgfplotsset{
height=6cm,
width=8.5cm,
}

\begin{document}

\title{Accelerating Innovation in 6G Research: \\Real-Time Capable SDR System Architecture\\ for Rapid Prototyping}

\author{
\IEEEauthorblockN{%
Maximilian~Engelhardt\IEEEauthorrefmark{2}\,\orcidlink{0009-0002-9440-8615},
Sebastian~Giehl\IEEEauthorrefmark{1}\,\orcidlink{0009-0008-1672-1351},
Michael~Schubert\IEEEauthorrefmark{2}\,\orcidlink{0009-0003-4843-2507},
Alexander~Ihlow\IEEEauthorrefmark{1}\IEEEauthorrefmark{2}\,\orcidlink{0000-0002-9714-4881},\\
Christian~Schneider\IEEEauthorrefmark{1}\IEEEauthorrefmark{2}\,\orcidlink{0000-0003-1833-4562},
Alexander~Ebert\IEEEauthorrefmark{1}\IEEEauthorrefmark{2}\,\orcidlink{0000-0001-6814-8173},
Markus~Landmann\IEEEauthorrefmark{2}\,\orcidlink{0009-0001-4270-0342},\\
Giovanni~Del~Galdo\IEEEauthorrefmark{1}\IEEEauthorrefmark{2}\,\orcidlink{0000-0002-7195-4253},
Carsten~Andrich\IEEEauthorrefmark{1}\,\orcidlink{0000-0002-4795-3517}
} \\[0.5ex]
\IEEEauthorblockA{\IEEEauthorrefmark{1}Institute of Information Technology, Technische Universität Ilmenau, Ilmenau, Germany}\\
\IEEEauthorblockA{\IEEEauthorrefmark{2}Fraunhofer Institute for Integrated Circuits IIS, Ilmenau, Germany}%
\thanks{
The authors acknowledge the financial support by the Bavarian Ministry of Economic Affairs, Regional Development and Energy in the project ``DSAI'' and by the Federal Ministry of Education and Research of Germany in the project ``6G-ICAS4Mobility'' (grant number: 16KISK241).
Furthermore, the research has been funded by the Federal State of Thuringia, Germany, and the European Social Fund (ESF) in the projects ``ML4ASP'' (grant number: 2019 FGI 0031) and ``MOTA'' (grant number: 2018 FGI 0041).
}
}

\maketitle

\begin{abstract}
The upcoming 3GPP global mobile communication standard 6G strives to push the technological limits of radio frequency (RF) communication even further than its predecessors:
Sum data rates beyond 100\,Gbit/s, RF bandwidths above 1\,GHz per link, and sub-millisecond latency necessitate very high performance development tools.
We propose a new SDR firmware and software architecture designed explicitly to meet these challenging requirements.
It relies on Ethernet and commercial off-the-shelf network and server components to maximize flexibility and to reduce costs.
We analyze state-of-the-art solutions (USRP X440 and other RFSoC-based systems), derive architectural design goals, explain resulting design decision in detail, and exemplify our architecture's implementation on the XCZU48DR RFSoC.
Finally, we validate its performance via measurements and outline how the architecture surpasses the state of the art with respect to sustained RF recording, while maintaining high Ethernet bandwidth efficiency.
Building a 6G integrated sensing and communication (ISAC) example, we demonstrate its real-time and rapid application development capabilities.
\end{abstract}


\section{Introduction}
Working towards the upcoming Release-19, the 3GPP has introduced \ac{ISAC} into the work plan for 5G NR~\cite{3gpp.5GNR.ISAC-WI}.
A feasibility study on the chances and requirements of \ac{ISAC} has been conducted beforehand~\cite{3gpp.5GNR.ISACcasestudy} and another one, targeting the relevance of channel modeling for \ac{ISAC}, is ongoing~\cite{3gpp.5GNR.ISAC-channelmodel}.

The ITU, contributing research organizations, and leading hardware vendors do agree, that \ac{ISAC} will also be part of the emerging 6G standard~\cite{ITU2022,ITU2023,rosemannJCAS6G,Flagship6GSensing,ericsson6g,huawei6g,nokia6g,nokia6g.2024,samsung6g,zte6g}.
Their use case visions define 6G's technological cornerstones:
Sum data rates in excess of \SI{100}{\giga\bit\per\second}, \ac{RF} bandwidths in excess of \SI{1}{\giga\hertz}, sub-millisecond latency, and possibly full-duplex radio communication.
The increased complexity and scope of 6G mandates a highly efficient research and development process to sustainably deliver the required results in time.~\cite{3gpp.6G.EU,3gpp.6G.US,Flagship6G-BW-UseCase}

In contrast to application-specific monolithic architectures, the \ac{SDR} concept~\cite{sdr_initial} provides the flexibility to cover a wide range of requirements, making it a popular choice in research and development.
However, the ease of use of existing ready-to-run \acp{SDR} in ecosystems such as GNU Radio or MATLAB is counterbalanced by limitations in terms of throughput and latency, which would, e.g., require individual adaptation in low-level system programming with C/\Cpp or even \ac{FPGA} design.
The performance requirements of 6G, in particular the high data rates and bandwidths paired with low latency, certainly necessitate the low-level programming approach.
Our own research endeavors of the last two decades~\cite{RUSK,danielSounding,standardLDM,EarlyICAS} reinforce our conviction that \ac{SDR} represents the best approach to mobile communication research and development.

In this contribution, we propose a novel \ac{SDR} system architecture that bridges the gap between support for rapid application development and sustained high performance, i.e., high throughput and low latency.
To achieve the required performance, we still have to rely on a highly parallel low-level \ac{FPGA} design and heavily optimized C/\Cpp software where necessary.
We establish zero-overhead interfaces for real-time signal processing modules.
All non-performance-critical parts, especially application-specific control code, are implemented as high-level scripting using a concurrent interface that provides deterministically timed access to the hardware functions, e.g., \ac{RF} frontends, sample acquisition, and waveform generation.

This architecture offers an unparalleled set of features compared to other \ac{sota} solutions~\cite{keysightdig,ADQ7DC,RFSoCRecorder,X410,X440,KITRFSoC1,KITRFSoC2}:
\begin{itemize}
  \item \textit{Strong isolation between scripted application logic and the hardware abstraction layer} resolves performance issues of \ac{sota} solutions where high-level scripts interfere with low-level realization.
  Additionally, realizing application logic in high-level scripts reduces the initial hurdle for developers and enables rapid application development.

  \item \textit{Full exploitation of the hardware's capabilities.}
  The high-speed serial link between \ac{FPGA} and host server is the system bottleneck.
  Therefore, the streaming efficiency in \ac{Rx} and \ac{Tx} is optimized towards the link's data rate limit.
  Beyond that, a separate solution (comparable to an arbitrary waveform generator) for periodic \ac{Tx} signals allows simultaneous use of all DACs at the maximum sample rate.

  \item \textit{Continuous sample recording only limited by the size of the SSDs.}
  Using hardware comparable to the \ac{sota}, the architecture realizes uninterrupted recording of a \SI{10}{\giga\byte\per\second} sample stream via a single \SI{100}{\giga\bit\per\second} Ethernet link.\footnote{This corresponds to a real-valued \SI{16}{\bit} sample signal at \SI{5}{\giga\sample\per\second} or an equivalent complex 2x\SI{16}{\bit} sample baseband signal at \SI{2.5}{\giga\sample\per\second}. Due to analog and digital filter roll-off, the modulated bandwidth of either signal is limited to approximately \SI{2}{\giga\hertz}.}
\end{itemize}

The remainder of the paper is structured as follows:
Starting from the state of the art in~\autoref{sec:SotA}, we derive the design goals that are required to establish such an architecture in~\autoref{sec:goals}.
Afterwards, we point out our design decisions and the resulting \ac{SDR} system architecture in Sec.~\ref{sec:SystemOV} to~\ref{sec:DAC}.
Finally, we portray our implementation on a Xilinx RFSoC and \ac{COTS} server hardware in~\autoref{sec:impl} and discuss the achieved performance based on system benchmarks in~\autoref{sec:txeval} and an exemplary 6G \gls{ISAC} measurement in~\autoref{sec:radar}.

\section{State of the Art}\label{sec:SotA}
\autoref{tab:SotA} enumerates a selection of architecture solutions and their realization.
The existing approaches can be split into two groups, namely dedicated monolithic architectures (1. and 2.) and classic \ac{SDR} systems (3. through 5.).

\begin{table*}[]
  \centering
    \caption{Comparison of the figures of merit of various wideband sampling solutions.}
    \label{tab:SotA}

      \begin{tabular}{r|l|S[table-format=4.2,table-number-alignment = center]|S|S|c}
        \toprule
          \textbf{Index}&\textbf{System} & {\textbf{Max. Instantaneous}} & {\textbf{Max. Rx}} & {\textbf{Data Rate}} & \textbf{Interface}\\
           && {\textbf{Bandwidth / \unit{\mega\hertz}}} & {\textbf{Channels}} & {\textbf{\,\unit{\giga\bit\per\second}\,}} & \\
        \midrule
        1. & Keysight Digitizer~\cite{keysightdig} & 12500 & 4 & 160 & Optical data interface (ODI)\\
        2. & Teledyne ADQ7DC~\cite{ADQ7DC} & 3000 & 2 & 56 & PCI Express\\
        3. & Universal RFSoC-based Signal Recorder~\cite{RFSoCRecorder} & 256 & 8 & 10& 10G Ethernet\\
        4. & Ettus USRP X410~\cite{X410} & 400 & 4 & 200 & 2x100G Ethernet\\
        5. & Ettus USRP X440~\cite{X440} & 1600 & 8 & 200 & 2x100G Ethernet\\
        6. & KIT MIMO testbed~\cite{KITRFSoC1,KITRFSoC2} & 2000 & 8 & 200 & 2x100G Ethernet\\
        \bottomrule
      \end{tabular}
\end{table*}

Monolithic solutions are meant to be composed of dedicated hardware framework components.
Architecture approaches like these usually -- due to the fine-tuned component selection -- reach outstanding performance, but lack flexibility due to hardware dependencies and proprietary data interfaces.
Adapting the system to different applications therefore requires changing its setup in components and low-level programming.
This results in prolonged development cycles and increased cost.

On the other hand, classic \ac{SDR} approaches gain flexibility by realizing a split architecture design~\cite{sdr_initial}.
They define a protocol layer, which realizes three planes: Control, \ac{Rx}, and \ac{Tx}.
The transceiver hardware must be compatible with the protocol; aside from that, it is interchangeable.
Using an appropriate data link, it is connected to a host \gls{PC}, where a software driver interacts with the transceiver hardware to realize generic functionality.
The driver offers a high-level interface to the user.
The actual application-specific code can be built on this interface and is therefore largely independent of the utilized transceiver hardware and vice versa.
In general, this enables designing a powerful and hardware independent architecture, which can utilize popular standard high-speed interfaces and \ac{COTS} hardware.
As trade-off against the increased flexibility, many \ac{SDR} architecture realizations suffer from performance limitations.

Solutions 3 to 5 utilize the Xilinx RFSoC chipset, which combines direct RF-sampling data converters with \ac{FPGA} hardware.
It offers eight \acp{ADC} and \acp{DAC} with up to \SI{5}{\giga\sample\per\second} each and two \SI{100}{\giga\bit\per\second} Ethernet interfaces.
Comparing the realized figures of merit, the impact of architecture design can easily be seen:
Solution 3 utilizes direct \ac{RF}-sampling, but also digital down conversion.
This allows for a possibly high channel count, but cuts down the realizable instantaneous bandwidth.
Furthermore, the underlying architecture only implements receiver operation.

In contrast, solutions 4 and 5 represent the two RFSoC-based \acp{SDR} available as \ac{COTS} devices from Ettus Research.
Both are ready to run solutions and include the necessary peripheral components and interconnects.
The Ettus open source software suite \gls{UHD} can be utilized for both devices.
The \ac{UHD} is accompanied by the \ac{RFNoC} architecture, which provides partially run-time reconfigurable hardware acceleration for signal processing by exploiting the computational capabilities of the \ac{FPGA}.
Although it simplifies the \ac{FPGA} design flow through its modular structure, it increases the overhead in terms of resource usage and data rate.
Design decisions required at compile-time limit the flexibility that can be achieved.
The X410 integrates an analog frontend for mixing and filtering which limits the bandwidth per channel.
In contrast, the X440 directly connects the balun coupled chassis inputs to the converters' pins and therefore supports enhanced bandwidths.
Ettus specifies the reference system to be only capable of utilizing the available two \SI{100}{\giga\bit\per\second} Ethernet links to up to 61\%.
The worst case configuration even reaches only 34\% of link utilization~\cite{X440kb}, which severely degrades performance.
Furthermore, as shown in~\cite{DanielUSRPSwitching}, although Ettus' driver allows to specify the execution time of commands, the architecture does not support parallel synchronous switching operations.
Moreover, the driver's interface is not designed for burst \ac{Tx} streaming, massively increasing the control overhead for the high-level script in applications such as \gls{MIMO} channel sounding.

To overcome the existing \ac{SDR} systems' data rate limitations, solution~6 is being developed in~\cite{KITRFSoC1,KITRFSoC2}.
The RFSoC-based \gls{MIMO} testbed presented there achieves outstanding performance, utilizing the links between host and converter device to over 90\%.
However, depending on the application, further optimization is desirable to reduce the number of samples transferred over these links:
In the \gls{Rx} path, burst sampling is not supported, and in the \gls{Tx} path, periodic signals have to be streamed continuously through the host.
Due to the targeted application, this system is not designed for continuous streaming and storage on SSDs, but buffers Rx samples in RAM during the acquisition, which limits the measurement duration to a maximum of \SI{10}{\second}~\cite{KITRFSoC2}.

\section{Design Goals}\label{sec:goals}

In order to design a platform as versatile as possible, the basic functionality should be isolated from the interchangeable application logic.
Similar to the \gls{UHD}, we aimed for a high-level application programming interface which allows to control all low-level functionalities from a programming language of the users's choice, such as Python.
The application logic can thus be modified without recompiling FPGA design or \Cpp code, also obviating the need to flash and to reinitialize the device.
This saves from a few minutes up to several hours per development iteration and is one central reason for the widespread success of \glspl{SDR} in research and development.
Beyond this, we aimed to achieve this isolation not only functionally, but also in terms of performance:
Decoupling of software components prevents blocking of critical tasks like sample recording by concurrent non-critical processes, e.g., real-time data analysis.
Asynchronous procedure calls allow commands to be issued without waiting for previous operations to complete.

We designed our architecture from the ground up for real-time capability:
This allows the application to react to measurement data with low latency and to precisely control the timing of all hardware interactions, which is essential in highly parallel \gls{MIMO} and multi-node setups.
One potential application is the implementation of an \acl{AGC} in software.

As a result, the following design objectives guided the development of our architecture:
\begin{enumerate}
 \item Optimize the data rate both over the host-to-device link and to the SSD storage to enable continuous recording
 \item Isolation between application logic and abstracted basic functionalities
 \item Shifting the complexity from the FPGA to the host, maxi\-mi\-zing the use of Python and minimizing the use of \Cpp
 \item Low latency in both control and data planes
 \item Deterministic timing of hardware actions independent of software timing jitter
 \item Flexibility in hardware, number of nodes/channels, and applications.
\end{enumerate}

These goals allow to create one versatile platform that can be adapted via high-level user scripts to a multitude of research and development wideband RF use cases, particularly in the context of 6G design, rapid prototyping, and implementation.
This encompasses applications such as multi-node MIMO RF propagation measurements, radar target characterization, \ac{ISAC}, and real-time algorithm testing, which also incorporates AI-driven features.

\section{System Architecture Overview}\label{sec:SystemOV}

\begin{figure*}[bt!]
\centering
\scalebox{0.8}{\begin{tikzpicture}

    \definecolor{color0}{rgb}{0.12156862745098,0.466666666666667,0.705882352941177}
\definecolor{color1}{rgb}{1,0.498039215686275,0.0549019607843137}
\definecolor{color2}{rgb}{0.172549019607843,0.627450980392157,0.172549019607843}
\definecolor{color3}{rgb}{0.83921568627451,0.152941176470588,0.156862745098039}
\definecolor{color4}{rgb}{0.580392156862745,0.403921568627451,0.741176470588235}
\definecolor{color5}{rgb}{0.549019607843137,0.337254901960784,0.294117647058824}

    \tikzset{
       boxa/.style={rectangle,draw=black, top color=white, inner sep=5pt,minimum width=1.2cm, minimum height=0.4cm, text centered, text width=2.5cm},
       arw/.style={>={Triangle[length=5mm,width=8mm]},line width=5mm,draw=lightgray},
       axi/.style={draw=color3}
    }

        \node[boxa] (axihost) {AXIoE Client};

        \node[boxa, below=0.5cm of axihost] (rxhost)  {Rx Path 1};
        \node[boxa, below=of rxhost] (txhost)  {Tx Path 1};
    \node[below=-0.2cm of rxhost] (rxhostdots) {\vdots};
    \node[below=-0.2cm of txhost] (txhostdots) {\vdots};

        \node[fit=(axihost)(rxhost)(txhost)(txhostdots), draw, black,  inner sep=0.2cm, rectangle,label={[name=dpdklabel]above:DPDK-Based Driver}] (dpdk) {};

    \coordinate (dpdk-p1) at ($(dpdk.north west)!0.25!(dpdk.south west)$);
    \coordinate (dpdk-p2) at ($(dpdk.north west)!0.75!(dpdk.south west)$);

    \node[boxa, left=of dpdk-p1] (py)  {Python Application};
    \node[boxa, left=of dpdk-p2] (ssd)  {Storage};

    \draw[<->] (py) -- node [midway,above,rotate=90,anchor=west,text width=4cm] {shared \\memory} (dpdk-p1);
    \draw[<->] (ssd) -- node [midway,above,rotate=90,anchor=west] {io\_uring} (dpdk-p2);

    \node[fit=(ssd)(dpdklabel)(dpdk), inner sep=0.5cm, draw, black, rectangle,label=above:Host] (host) {};

    \node at ($(axihost.east)+(4.75,.5)+(0,-0.1pt)$) [boxa] (axifpga) {AXIoE Server};
    \node[boxa, below=0.5cm of axifpga] (rxfpga) {Rx Path 1};
    \node[boxa, below=of rxfpga] (txfpga) {Tx Path 1};
    \node[below=-0.2cm of rxfpga] (rxfpgadots) {\vdots};
    \node[below=-0.2cm of txfpga] (txfpgadots) {\vdots};
    \node[boxa, below=of txfpga] (periph) {Peripheral Control};

    \node[right=1.9cm of rxfpga] (adc1) {ADC 1};
    \node[right=1.9cm of txfpga] (dac1) {DAC 1};
    \node[below=-0.2cm of adc1] (rxadcdots) {\vdots};
    \node[below=-0.2cm of dac1] (txdacdots) {\vdots};

    \node[fit=(adc1)(txdacdots), inner sep=0.5cm, draw, black, rectangle,label={above:Data Converters}] (rfdc) {};

    \node[boxa,anchor=south east] at (periph.south -| rfdc.east) (clocking) {Clocking};

    \node[fit=(axifpga)(txfpgadots)(clocking)(periph), inner sep=0.7cm, draw, black, rectangle,label=above:Converter Device FPGA] (fpga) {};

    \draw[arw,<->] (host.east) -- (host.east-|fpga.west);
    \coordinate (eth-left) at ([xshift=-0.25cm]host.east);
    \coordinate (eth-right) at ([xshift=0.25cm]host.east-|fpga.west);

    \node[text width=1.8cm, text centered] at ([yshift=0.8cm]$(host.east)!0.5!(host.east-|fpga.west)$) {$\geq\!\!100$\,Gbit/s Ethernet};

    \draw[color0, <->] (axihost.east) -- ([xshift=0.25cm]axihost.east) -- ([yshift=0.1cm]eth-left) -- ([yshift=0.1cm]eth-right) -- ([xshift=-0.25cm]axifpga.west) -- (axifpga.west);
    \draw[color1, <-]  (rxhost.east) -- ([xshift=0.25cm]rxhost.east) -- (eth-left) -- (eth-right) -- ([xshift=-0.25cm]rxfpga.west) -- (rxfpga.west);
    \draw[color2, ->]  (txhost.east) -- ([xshift=0.25cm]txhost.east) -- ([yshift=-0.1cm]eth-left) -- ([yshift=-0.1cm]eth-right) -- ([xshift=-0.25cm]txfpga.west) -- (txfpga.west);

    \draw[<-] (rxfpga) -- (adc1);
    \draw[->] (txfpga) -- (dac1);

    \draw[<-] (adc1.east) -- node[pos=1, right] {Rx Signal} ([xshift=1.5cm]adc1.east);
    \draw[->] (dac1.east) -- node[pos=1, right] {Tx Signal} ([xshift=1.5cm]dac1.east);

    \draw[axi,<->] (axifpga.east) -- ([xshift=0.7cm]axifpga.east) |- ([yshift=0.3cm]clocking.north west)-- ([xshift= 0.3cm]clocking.north west);
    \draw[axi,->] (axifpga.east) -- ([xshift=0.7cm]axifpga.east) |- ([yshift=0.3cm]rxfpga.north east) -- ([xshift=-0.3cm]rxfpga.north east);
    \draw[axi,->] (axifpga.east) -- ([xshift=0.7cm]axifpga.east) |- ([yshift=1cm]rfdc.south west);
    \draw[axi,->] (axifpga.east) -- ([xshift=0.7cm]axifpga.east) |- node[color=color3, midway, below, rotate=90,xshift=0.2cm] (axi) {AXI4 Bus} ([yshift=0.3cm]txfpga.north east) -- ([xshift=-0.3cm]txfpga.north east);
    \draw[axi,->] (axifpga.east) -- ([xshift=0.7cm]axifpga.east) |- ([yshift=0.3cm]periph.north east) -- ([xshift=-0.3cm]periph.north east);

    \coordinate (periph1) at ([xshift=-0.4cm]periph.south);
    \coordinate (periph2) at ([xshift=0.4cm]periph.south);
    \draw[->] (periph1) -- node[pos=1, below] {I$^2$C} ([yshift=-1cm]periph1);
    \draw[->] (periph2) -- node[pos=1, below] {\phantom{$^2$}SPI\phantom{$^2$}} ([yshift=-1cm]periph2);

    \coordinate (clk1) at ([xshift=-0.8cm]clocking.south);
    \coordinate (clk2) at ([xshift=0.8cm]clocking.south);
    \draw[<-] (clk1) -- node[pos=1, below] {\phantom{$^2$}REFCLK\phantom{$^2$}} ([yshift=-1cm]clk1);
    \draw[<-] (clk2) -- node[pos=1, below] {\phantom{$^2$}SYSREF\phantom{$^2$}} ([yshift=-1cm]clk2);
\end{tikzpicture}}
\caption{
Overview of our system architecture, consisting of the converter device, which is operated by a driver software on the host computer via one or multiple high-speed Ethernet links. Its generic design allows for adaption and exchange of hardware components and high-speed data interface. This allows to achieve the individually desired performance parameters without requiring changes to the architecture. Encapsulation of base features in an abstraction layer on the host allows for interfacing with the system via the high-level scripting language Python.}
\label{fig:SystemOverview}
\end{figure*}
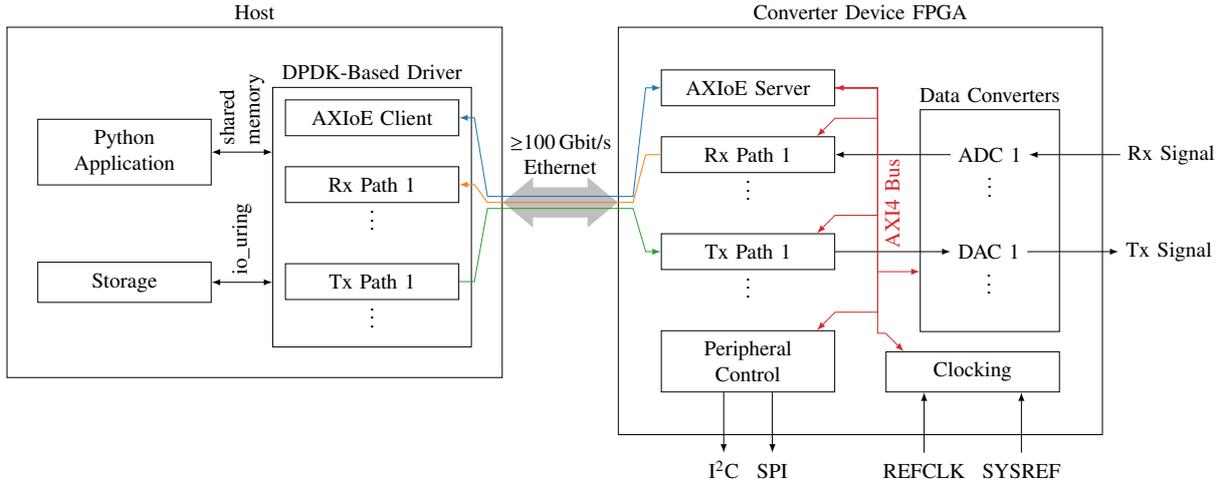

\autoref{fig:SystemOverview} outlines the basic structure of our proposed architecture, which consists of the PC-based host and the converter device, with an \ac{FPGA} at its core, connected via a high-speed Ethernet link.
Due to its generic design, the hardware components as well as the high-speed data interface can be exchanged to match individual requirements without incurring changes to the architecture.\footnote{Minor changes to the implementation might be required.}

Ethernet provides the most promising solution to connect host and converter device, as suitable \glspl{NIC} are widely available for \ac{COTS} server hardware.
It supports high data rates and allows for a wide range of connection types:
From short direct connections via copper cable to kilometer-long fiber optic cables and complex switched networks.
Using Ethernet as an interchangeable, generic, and flexible interface, our architecture overcomes the limitations of monolithic \ac{sota} setups and proprietary interfaces.

The Ethernet interface is used for both the communication to control the device (control plane) and the streaming of samples (data plane).
For the latter, there is one data path per channel, whereby a distinction is made between \ac{Rx} (\ac{ADC} channels, sample flow from device to host) and \ac{Tx} (\ac{DAC} channels, reverse sample flow).
The planes' individual communication protocols were each designed with the goals of efficiency and shifting complexity from the \ac{FPGA} to the host to simplify implementation.
Unlike \ac{UHD}, our architecture natively supports burst \ac{Tx} streaming, eliminating any control overhead.
Moreover, it is optimized for ultra-fast and overhead-free error recovery in both \ac{Rx} and \ac{Tx} streaming operation.

To link the control plane modules within the \ac{FPGA}, we employ the de facto standard \ac{AXI}.
All modules provide their configuration via registers, which can be read and written through this memory-mapped interface.
As \ac{AXI} is widely used, a variety of ready-made infrastructure cores and functional IP cores with compliant control ports are available.
Bridges from the \ac{AXI} network to other protocols, e.g., \ac{SPI} or \ac{I2C}, integrate peripheral components like analog \gls{RF} frontends.
The \ac{AXIoE} protocol~\cite{AXIoEPaper,AXIoESpec} tunnels register accesses through Ethernet and enables direct access to all components via straightforward Python\footnote{Without loss of generality, we utilize Python to implement the application logic due to its widespread use, readability, and extensive library support.} programming on the host.
On the FPGA, a timed command infrastructure enables the precise timing of actions independent of software timing jitter.
Combined, this facilitates rapid prototyping with analog components, e.g., development of advanced 6G features like full-duplex \gls{RF} transceivers.

For all data streams within the \ac{FPGA}, e.g., \gls{ADC} and \gls{DAC} samples, the unidirectional variant \ac{AXI}-Stream is used.

The link to the host is the bottleneck of the architecture as the combined data rate of the utilized converter channels may exceed its capacity.
In that case, not all streams can be continuously operated with maximum \ac{RF} bandwidth and 100\% duty cycle.

To realize maximum flexibility, each channel features an independent data path on the \ac{FPGA}, which allows time-based sampling control per channel, supporting both burst and continuous sampling modes.
Samples from individual channels are independently packed into Ethernet packets, which are then transmitted on the Ethernet interface in a round-robin arbitration scheme.
All Ethernet packet headers include \ac{VLAN} tags to distinguish between channels and to enable building distributed setups with multiple converter and host nodes connected to a switched network.
To ensure only intact packets are processed, the \ac{FPGA} monitors all inbound packets' \ac{CRC} and \ac{RS-FEC} status.
Faulty packets will be discarded.

The host-side software is separated into two components:

\begin{enumerate}
\item A driver which encloses the base functionality in low-level code and the high-level application-specific scripting.
The driver, written in \Cpp, handles high-rate and latency-critical communication.
This is supported by the \gls{DPDK} framework, which is a widely used solution for performance-optimized networking in the data center industry~\cite{DPDK}.
Like the \ac{FPGA} image, the driver is developed to be generic and reusable.
\item A Python program governs the application-specific operations.
This program interacts with the driver via a shared memory interface, implementing mechanisms for both asynchronous \ac{RPC} and sample transfer.
This separation allows the application logic to be modified without recompiling \Cpp code or even the FPGA design, also obviating the need to flash and reinitialize the SDR device.
\end{enumerate}

We realize continuous streaming and recording of a \SI{10}{\giga\byte\per\second} sample stream via a single \SI{100}{\giga\bit\per\second} Ethernet link.
Thus, we overcome the data rate and duty cycle limitations of \ac{UHD}~\cite{X440} and the \ac{MIMO} testbed presented in~\cite{KITRFSoC1,KITRFSoC2}.

\section{Synchronization \& Timing}\label{sec:Sync}
In line with our design goals, we realize synchronous sampling and deterministically timed command execution in distributed setups.
To allow for synchronous sampling, a \ac{REFCLK} is distributed to each converter device.
Absolute synchronizability in time is gained by additionally introducing a \ac{SYSREF} and using a simplified synchronization scheme similar to JESD204~\cite{JESD204D}.
It is generated and distributed with a defined timing relation to the \ac{REFCLK}.
All clocks required internally are derived from the \ac{REFCLK} on each individual converter device.
Ensuring integer clock relations yields deterministic synchronizability as well as synchronous sampling for local and distributed converter device setups.
We plan to publish a more detailed description of the clock generation, distribution, and synchronization scheme for single and multi-device setups in a dedicated article.

In order to arrange commands and data in a chronological sequence, two time bases are employed.
Firstly, we define a sampling-oriented time base:
The data converters aggregate samples into beats resulting in a configuration-dependent beatclock.
From this, we derive the \textbf{beatcounter} which provides a direct relation between sample beats and time.
It is used for all sampling-related purposes.
Secondly, as low sample rates unavoidably incur insufficient resolution of the beatcounter, we derive an invariant \textbf{real time clock (RTC)}\glsunset{RTC} from the \ac{FPGA}'s \ac{REFCLK}.
It is used for all other timing needs, e.g., interaction with peripheral components.

\subsection{FPGA Sync and Time}
The \ac{FPGA} implements individual counters for both time bases.
Since the \ac{RTC} is clocked by the \ac{REFCLK}, it can be directly synchronized using the external \ac{SYSREF} signal, which has a known timing relation to the \ac{REFCLK}.
The synchronization mechanism is armed via \ac{AXIoE} and will subsequently trigger on the \ac{SYSREF}'s next rising edge.
Based on the previously synchronized \ac{RTC}, the beatcounter is initialized via a timed command.
The synchronization algorithm predicts both clocks' common edges and thereby ensures deterministic operation for all possible (integer) \ac{REFCLK} to beatclock relations no matter which possesses the highest frequency.

The implementation of these fast counters is a major challenge.
In fact, their adder's possibly long carry chain might impair the design's timing closure and therefore requires designing ultra-fast and lightweight counter modules.
Instead of implementing a monolithic adder structure, the logic is split into two synchronized counters as proposed by~\cite{WilliPaper}.
While the first counter increments the lower part of the output word in every cycle, the second one precalculates the next upper part and propagates it to the output as soon as the fast counter overflows.
This way, the long carry chain can be split into two parts.
While the first counter's short chain still has to fulfill the tight requirements, the second counter's long chain's timing can be eased by applying multicycle path constraints.\footnote{The generic design allows for compile-time reconfiguration of the counter's split.}

\subsection{Host-side Packet Pacing}\label{pacing}
Most \acp{FPGA} have very limited internal buffer capabilities, e.g., in the order of a few megabits.
Without incurring further hardware dependencies like utilization of external memory, the SDR may only buffer a few microseconds of \ac{Tx} sample data.
To prevent overflows, the host must not send its samples to the device too early.
Similarly, on the control plane, the host must adhere to queue size limitations of the \ac{FPGA}.
Therefore, a flow control mechanism is required.

In the \textit{\ac{CHDR}} protocol used by the \acp{USRP}, the device gives clearances to the host to transmit data up to a given stream position (transmit window)~\cite{RFNoCSpec}.
As it relies on the host to react in time, the buffers in the \ac{FPGA} must be significantly larger than the worst-case software latency.
Therefore, this solution is not viable here.
In the newer \ac{USRP} X440, Ethernet pause frames are used for flow control~\cite{X440kb}.
These are packets that request the remote device to suspend data transmission for a specified period of time.
However, the disadvantage of this method is that it is not channel specific and, in particular, slows down time-critical control communication (head-of-line blocking).

Instead, we take a different approach:
For both timed commands and \ac{Tx} samples, it is known when the device will consume them from its buffers.
This allows the host to send the packets accurately timed so that they arrive at the device with a fixed lead time.
To achieve this precisely, we employ the \textit{send on timestamp} feature of the \ac{NIC} hardware, which allows the software to schedule packets ahead of time~\cite{DPDK_NIC}.
The \gls{NIC} puts the packets on the wire at the predetermined transmission time.
Thus, the arrival time at the \ac{FPGA} is decoupled from the relatively high software timing jitter, rendering significantly smaller buffers possible.

To achieve this, the host has to know the relationship between its local and the device's remote time bases.
In fixed intervals, it requests the device to read out the current time stamp counter value.
The response then tells the time at which the request packet arrived at the device.
This is used as input to a clock model to adjust the transmission timestamps of future packets.

\section{Device Control}\label{sec:Control}
Full-featured remote control of the \ac{FPGA}'s internal modules and peripheral components by the host is required to enable an architecture independent from both hardware and application.
In the \ac{FPGA} design, we realize all configuration and control functionality via memory-mapped \ac{AXI}.
We utilize \ac{AXIoE} as generic and reliable access protocol for tunneling the \ac{AXI} accesses from the host to the \ac{FPGA} over Ethernet.
This allows for easy integration of commercial and custom modules and peripheral components into the architecture.

\subsection{AXI over Ethernet}
For the remote control of the different blocks on the \ac{FPGA}, we employ the protocol \ac{AXIoE}, specified in~\cite{AXIoESpec}, which allows to tunnel memory-mapped \ac{AXI} accesses via Ethernet~\cite{AXIoEPaper}.

The protocol realizes an ordered and reliable command stream, but also allows for independent commands to be transmitted unreliably and out of sequence.
It is specifically designed for the constraints of the communication between a host and an \ac{FPGA}:
Its asymmetric design requires the \ac{AXIoE} server on the \ac{FPGA} side to only realize a simple request-response mechanism.
In contrast, the error detection and recovery process is to be realized by the client running on the host.
This allows for a lightweight \ac{FPGA} implementation, shifting complexity to the host software.

The host can send request packets, which may contain one or multiple transactions.
Each transaction consists of one atomic command, either a read from or write to a specified memory-mapped address range.
Upon packet loss, the host performs a resynchronization:
It either repeats the lost request packets or asks the \ac{FPGA} to repeat lost responses.

\begin{figure}[bt!]
  \centering
  \scalebox{0.7}{\begin{tikzpicture}

    \definecolor{color0}{rgb}{0.12156862745098,0.466666666666667,0.705882352941177}
    \definecolor{color1}{rgb}{1,0.498039215686275,0.0549019607843137}
    \definecolor{color2}{rgb}{0.172549019607843,0.627450980392157,0.172549019607843}
    \definecolor{color3}{rgb}{0.83921568627451,0.152941176470588,0.156862745098039}
    \definecolor{color4}{rgb}{0.580392156862745,0.403921568627451,0.741176470588235}
    \definecolor{color5}{rgb}{0.549019607843137,0.337254901960784,0.294117647058824}

    \tikzset{
       boxa/.style={rectangle, draw =black,minimum width = 2cm, minimum height = 1cm,align = center}
    }

    \node [boxa](Preproc){AXIoE\\Preprocessing};
    \node [right = .5cm of Preproc] [boxa](CmdF){Command\\FIFO};
    \node [right = .5cm of CmdF] [boxa, minimum height = 1.5cm](Machine){AXIoE\\State\\Machine};
    \node [above = .5cm of Machine] [boxa](RespBuf){Response\\FIFO};
    \node [below = .5cm of Machine] [boxa](RetrBuf) {Replay\\Memory};

    \node at ($(Machine.east) + (.3,.5)$)(align0){};
    \node at ($(Machine.west) + (-.3,.5)$)(align1){};

    \draw[->,line width = 1pt](Preproc) -- (CmdF);
    \draw[->,line width = 1pt](CmdF) -- (Machine);

    \draw[->,line width = 1pt](Machine.east|-align0.center) -- (align0.center) -- (align0.center|-RespBuf) -- (RespBuf);
    \draw[<-,line width = 1pt](Machine.west|-align1.center) -- (align1.center) -- (align1.center|-RespBuf) -- (RespBuf);

    \draw[->,line width = 1pt](Machine.east) -- ++( .5,0) node [right, align=center] {To Ethernet};
    \draw[<-,line width = 1pt](Preproc.west) -- ++(-.5,0) node [left, align=center] {From\\Ethernet};

    \draw[->,line width = 1pt](Machine.280) -- (Machine.280|-RetrBuf.north);
    \draw[<-,line width = 1pt](Machine.260) -- (Machine.260|-RetrBuf.north);

    \draw[color3,<->,line width = 1pt]($(Machine.east)+(0,-.5)$) -- ++(.5,0) node [color3, right] {AXI4 Bus};

\end{tikzpicture}}
  \caption{
    \ac{AXIoE} packet processing \ac{FPGA} design: \ac{AXI}-Stream-based command processing and \ac{AXI} bus master interface. Separate AXIoE preprocessing enables fast error recovery.}
  \label{fig:FPGAAXIoE}
\end{figure}
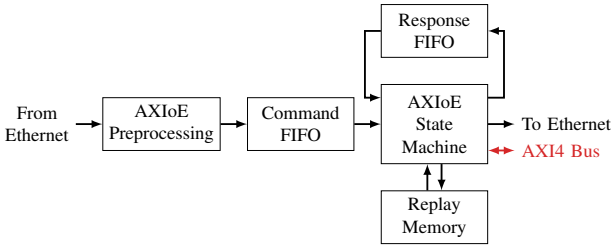

On the \ac{FPGA} side, we choose a split implementation of the server functionality as visualized in \autoref{fig:FPGAAXIoE}.
The preprocessing module checks incoming packets' \ac{AXIoE} headers.
Whereas protocol compliant packets are forwarded into the inbound command \ac{FIFO} buffer, faulty packets are discarded and error tickets -- containing the information required for generating the error response -- are inserted into the \ac{FIFO} instead.\footnote{This prevents faulty packets from entering the \ac{FIFO} and therefore fully eliminates the time required to read them out of the command \ac{FIFO} in error case.}
The \ac{AXIoE} state machine processes the incoming requests from the command \ac{FIFO}, executes transactions on the \ac{AXI} interface, and generates response packets.
For proper operation it requires utilizing two additional buffers:
The response \ac{FIFO} stores individual \ac{AXI} transaction response data until its status header can be generated.
The replay memory is addressable and stores full response pakets for potential replay requests.

\subsection{\glsfmtshort{FPGA} Command Timing}
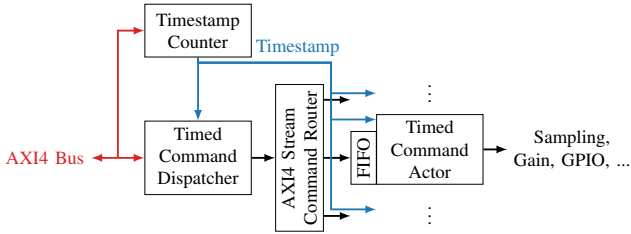
\begin{figure}[bt!]
  \centering
  \scalebox{0.7}{\begin{tikzpicture}

    \definecolor{color0}{rgb}{0.12156862745098,0.466666666666667,0.705882352941177}
    \definecolor{color1}{rgb}{1,0.498039215686275,0.0549019607843137}
    \definecolor{color2}{rgb}{0.172549019607843,0.627450980392157,0.172549019607843}
    \definecolor{color3}{rgb}{0.83921568627451,0.152941176470588,0.156862745098039}
    \definecolor{color4}{rgb}{0.580392156862745,0.403921568627451,0.741176470588235}
    \definecolor{color5}{rgb}{0.549019607843137,0.337254901960784,0.294117647058824}

    \tikzset{
       boxa/.style={rectangle, draw =black,minimum width = 2cm, minimum height = 1cm,align = center}
    }

    \node [boxa](Disp){Timed\\Command\\Dispatcher};
    \node [above = 1.2cm of Disp] [boxa](TSC){Timestamp\\Counter};				
    \node at ($(Disp.east)+(.9,0)$) [anchor=center,rectangle, draw =black,minimum width = .1cm, minimum height = .1cm, align = center, rotate = 90](Router){AXI4 Stream\\Command Router};
    \node at ($(Router.south)+(.75,0)$) [anchor = center,rectangle, draw =black,minimum width = 1cm, minimum height = .1cm, align = center, rotate = 90](fifo){FIFO};
    \node at ($(fifo.south west) + (-.25pt,0)$) [boxa, anchor= south west, minimum height = 1.25cm](act) {Timed\\Command\\Actor};
    \node at ($(act.east) + (.45,0)$) [align=center, anchor= west, minimum height = 1.25cm](periph) {Sampling,\\Gain, GPIO, ...};
    
    \node at ($(Router.south)+(0,1.1)$) (align0) {};
    \node at ($(Router.south)+(0,-1.1)$) (align1) {};
    \node at (fifo.north|-align0.center) (align2) {};
    \node at (fifo.north|-align1.center) (align3) {};

    \draw[->,line width = 1pt](Disp) -- (Router);
    \draw[->,line width = 1pt](Router) -- (fifo);
    \draw[->,line width = 1pt]($(Router.south)+(0,1.1)$) -- (align2.center);
    \draw[->,line width = 1pt]($(Router.south)+(0,-1.1)$) -- (align3.center);
    \draw[->,line width = 1pt](act) -- (periph);
    
    \node[below =0.1cm of act] {\vdots};
    \node[above =0.1cm of act] {\vdots};
    
    \draw[color3,<->,line width = 1pt](Disp.west) -- ++(-1,0) node [color3, left] (bus){AXI4 Bus};
    \draw[color3,->,line width = 1pt](bus.east) ++(.5,0) |- (TSC);

    \draw[color0,->,line width = 1pt](TSC) -- (Disp);
    \draw[color0,->,line width = 1pt](TSC.south) -- ++(0,-0.1) -- ++(2.5,0) node[above , near end] {Timestamp} |- ($(act.north west)+(0,-.1)$);
    \draw[color0,->,line width = 1pt](TSC.south) -- ++(0,-0.1) -- ++(2.5,0) |- (act.north west|-align2.north);
    \draw[color0,->,line width = 1pt](TSC.south) -- ++(0,-0.1) -- ++(2.5,0) |- (act.north west|-align3.north);

\end{tikzpicture}}
  \caption{
  Timed command network: \ac{FPGA} design topology. \ac{AXI} bus to \ac{AXI}-Stream conversion, \ac{AXI}-Stream routing, and timestamp distribution. Standard interfaces and lightweight design allow for simple and fast functional extension. Either beatcounter or RTC may be used as timestamp.}
  \label{fig:FPGAtcmd}
\end{figure}

\autoref{fig:FPGAtcmd} depicts the timed command network topology implemented on the \ac{FPGA}.
The respective time bases' counter output is distributed to all related blocks.
The timed command dispatcher module is accessible via the \ac{FPGA}'s \ac{AXI} network and converts memory-mapped accesses to \ac{AXI}-Stream packets containing the timed commands.
It checks the command time margin, generates the routing information, and inserts it alongside the actual packet data into the timed command \ac{AXI}-Stream network.
Each timed command actor features its own \gls{FIFO} buffer.
This allows queueing timed commands independently for each actor and avoids head-of-line blocking.
An actor module loads a command and executes it as soon as the execution target time is reached.
Due to the timing mechanism's time base independent design, each actor module may utilize either the beatcounter or the \ac{RTC}.

To further relax critical timing paths, the execution target time check is uncoupled from the actor module's actual command logic by setting a start bit and delaying the execution by one cycle instead of performing the check and the command in the same clock cycle.

\section{High-rate \glsfmtshort{ADC} Streaming}\label{sec:ADC}

When host and \ac{SDR} are connected via one \SI{100}{\giga\bit\per\second} Ethernet link, theoretically up to \SI{99.623}{\giga\bit\per\second} net data rate can be achieved when using jumbo frames of 9000 bytes payload.
We were able to demonstrate a data rate of \SI{95.885}{\giga\bit\per\second} for the maximum standard-compliant packet size of 1500 bytes.~\cite{ba_giehl}

To allow for any meaningful error recovery, the converter device would need to buffer the outbound data stream for a significant amount of time.
The internal memory resources typically found on \glspl{FPGA} do not suffice to realize this.
Using two or more DDR4 memory banks would provide the bandwidth required for prolonged buffering, but would introduce unwanted hardware dependencies on the system architecture.

The same reason also opposes the use of the \gls{RoCE} protocol, which is a common solution for implementing reliable high-rate data streams:
On the host side, it is directly supported by many \glspl{NIC}, allowing zero-copy interaction with continuous memory regions without CPU intervention.
However, it demands significant resources on the FPGA side~\cite{roce}.
Beyond this, \ac{RoCE}'s real-time capabilities are limited as the software on the host does not have control over the timing of individual packets.

Instead, we designed an ultralight \ac{Rx} protocol without retransmission capabilities.
Similarly to the control plane protocol, it is constructed asymmetrically to attain minimal implementation complexity on the \ac{FPGA}.
Thus, we centered our design around the sample beats generated by the \acp{ADC}:
They form atomic units, which will not be fragmented on the \ac{FPGA} and their generation frequency is the basis for their timestamping (see beatcounter in \autoref{sec:Sync}).

In addition to the actual sample data, each \ac{Rx} packet contains the following meta information:
\begin{itemize}
  \item The beatcounter based timestamp of the packet's first sample, allowing the host to infer the timestamp of all samples within the packet
  \item Packet sequence number, independent per channel
  \item \ac{ADC} status bits providing overrange, overvoltage, and threshold information, allowing the host to detect analog faults and implement an \acl{AGC}.
\end{itemize}

\subsection{FPGA-Design Realization}

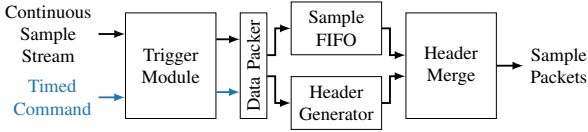
\begin{figure}[bt!]
  \centering
  \scalebox{0.7}{\begin{tikzpicture}

    \definecolor{color0}{rgb}{0.12156862745098,0.466666666666667,0.705882352941177}
    \definecolor{color1}{rgb}{1,0.498039215686275,0.0549019607843137}
    \definecolor{color2}{rgb}{0.172549019607843,0.627450980392157,0.172549019607843}
    \definecolor{color3}{rgb}{0.83921568627451,0.152941176470588,0.156862745098039}
    \definecolor{color4}{rgb}{0.580392156862745,0.403921568627451,0.741176470588235}
    \definecolor{color5}{rgb}{0.549019607843137,0.337254901960784,0.294117647058824}

    \tikzset{
       boxa/.style={rectangle, draw =black,minimum width = 1.7cm, minimum height = 1cm,align = center}
    }

    \node [boxa, minimum height = 2cm](TM){Trigger\\Module};
    \node [right = 0.7cm of TM] [boxa, minimum width = 2cm, minimum height = .5cm, anchor = center, rotate = 90](Pack){Data Packer};
    \node [right = 2.6cm of Pack.south] [boxa, minimum height = 2cm](HeaderMerge){Header\\Merge};
    \node [above = 0.2cm of $(Pack.south)!0.5!(HeaderMerge.west)$] [boxa](PackFIFO){Sample\\FIFO};
    \node [below = 0.2cm of $(Pack.south)!0.5!(HeaderMerge.west)$] [boxa](HeaderGen){Header\\Generator};
    
    \draw[->,line width = 1pt](HeaderMerge.east) -- ++(0.5,0)node [right, align=center]{Sample\\Packets};
    \draw[<-,line width = 1pt]($(TM.west)+(0,0.6)$) -- ++(-0.5,0)node [left, align=center]{Continuous\\Sample\\Stream};

    \draw[->,line width = 1pt]($(TM.east)+(0,0.5)$) -- ($(Pack.north)+(0,0.5)$);
    \draw[color0,->,line width = 1pt]($(TM.east)+(0,-0.5)$) -- ($(Pack.north)+(0,-0.5)$);
  
    \draw[->,line width = 1pt]($(Pack.south)+(0,+0.2)$) -- ++(0.15,0) |- (PackFIFO);
    \draw[->,line width = 1pt](PackFIFO.east) -- ++(0.15,0) |- ($(HeaderMerge.west)+(0,0.2)$);
    \draw[->,line width = 1pt]($(Pack.south)+(0,-0.2)$) -- ++(0.15,0) |- (HeaderGen);
    \draw[->,line width = 1pt](HeaderGen.east) -- ++(0.15,0) |- ($(HeaderMerge.west)+(0,-0.2)$);

    \draw[color0,<-,line width = 1pt]($(TM.west)+(0,-0.6)$) -- ++(-0.5,0) node[left, align = center]{Timed\\Command};
    
\end{tikzpicture}}
  \caption{
    \ac{Rx} \ac{FPGA} path: \ac{AXI}-Stream-based, beatcounter-timed triggering, MTU compliant packing, efficient buffering, parallel header generation, and data unit relation based merging. The design enables burst and continuous sample streaming.}
  \label{fig:FPGARxPath}
\end{figure}

\autoref{fig:FPGARxPath} illustrates the \ac{FPGA} side \ac{ADC} path design.
The trigger module executes beatcounter-timed commands, allowing for both burst or continuous sampling.
This module forwards an \ac{ADC} channel's sample beats as one \ac{AXI}-Stream packet per trigger event.
The first beat's beatcounter value is attached to the packet.
The subsequent module, i.e., the data packer, fragments the original packet according to the high-speed interface's \ac{MTU} size.
To maintain the stream's embedded timing information, intermediate beatcounter values are calculated internally and attached to each fragment.
The data packer directly feeds the sample data into the sample \ac{FIFO} for subsequent merging with the associated packet headers:
In parallel, the header generator produces one packet header per fragment from the accumulated metadata.
The header merge module combines the sample fragments with their respective headers and forwards them to the Ethernet interface.

In burst mode, the resulting data rates can exceed the capacity of the Ethernet link in the short term.
Therefore, buffering is necessary to cover the occurring backpressure.
Due to the limited resources of the \ac{FPGA}, an efficient design is essential.
This is achieved by splitting the data path into multiple, parallel sub-paths, e.g., for sample data and packet headers.
Each individual sub-path handles either data units of \ac{AXI}-Stream packets or single \ac{AXI}-Stream beats.
This relation empowers implicit synchronization by \ac{AXI}-Stream flow control and path merging by a lightweight mechanism based on the data units.

The data relation between the sub-paths enables a pipelineable, modular, and expandable path design and individual buffering or processing per path without explicitly implementing any synchronization mechanism.
To preserve this inter-path data relation, all modules within the paths must be designed to be fully flow control compliant.
A recovery mechanism handles backpressure conditions impacting the trigger module\footnote{To maintain the timing relation to the continuous stream of \ac{ADC} samples, it must not support backpressure.} and prevents the sub-paths from losing synchronization.

Special considerations are necessary to allow continuous operation of the \ac{ADC} paths.
Backpressure occurring in paths with critical load will eventually cause \ac{FIFO} overflows.
To prevent systematic backpressure, all modules used in paths with critical load must be implemented with an initiation interval\footnote{The delay between processing of successive input data in units of clock cycles.} of 1.
Stochastic backpressure must not occur either, since it can never be caught up with.
This requires an adequate path design with respect to data width conversions and clock domain crossings.

\subsection{Host-side Implementation}\label{sec:HostRecording}

A central use-case of our SDR architecture is recording the received samples for offline processing, requiring particularly fast access to large amounts of storage.
In order to achieve the maximum possible write rates, we use multiple SSDs and combine their individual write speeds using a software RAID0 and the XFS filesystem.
The host software interacts with the storage using \textit{io\_uring}, an asynchronous, high-performance interface to the Linux kernel~\cite{iouring}.

For maximum performance, we use the \textit{O\_DIRECT} access mode of the Linux kernel.
It imposes a fixed block size, in our case 512 bytes~\cite{odirect}.
This is in conflict with the requirements of the communication protocol, which handles sample beats as atomic units.
Therefore, we decided not to strive for a zero-copy implementation in the driver, but rather to copy the samples from the individual packets into a large ring buffer.
This provides maximum flexibility.
In fact, samples can be selected arbitrarily in this step without being bound to beat limits.
If samples are missing in the output stream, e.g., due to a lost packet, they are zero-padded to ensure that the subsequent samples are found at the expected position in the stream.
The ring buffer allows for simultaneous online processing of samples, e.g., to provide a live view of the data or to implement an \acl{AGC}, without disrupting the real-time recording in any way.

It would be a desirable feature to be able to configure the network card not to drop packets with erroneous checksums:
Since there is no retransmission option in our application, we would rather accept bit errors in the sample stream than lose entire packets.
Unfortunately, DPDK currently does not provide a way to configure the \gls{NIC} accordingly.

\section{Analog Signal Generation: \glsfmtshort{DAC}-Path}\label{sec:DAC}
In common SDR solutions, only streaming is supported in Tx mode, making the Ethernet link the bottleneck of the system.
To overcome this, we have implemented approaches to efficiently support both static and dynamic Tx sequences:

The first approach operates in a \textbf{loading-looping} manner -- like an arbitrary waveform generator -- realizing dynamic exchangeability of sequences at runtime via \ac{AXIoE}.
Since a sequence only needs to be transmitted to the converter device once, this mode of operation does not require high performance host hardware to setup and control the waveform generation.
It enables the synchronized playback of periodic sequences on multiple channels, vastly exceeding the Ethernet interface data rate.

The second approach implements the most flexible solution realizable for an \ac{SDR} platform: \textbf{Real-time \ac{Tx} sample streaming}.
The ability to transmit arbitrary RF signals enables the platform to implement a full-featured Tx.
In this setup, the host server transmits the \ac{DAC} samples and their associated timing information to the converter device, which buffers the data until the specified playback time.
To ensure proper sample handling, both ends implement mechanisms to compensate for network jitter.
If there are no samples to be played, zero-samples are automatically passed to the respective \acp{DAC} by the FPGA (auto-zero on idle).
In contrast to \ac{UHD}, this eliminates the need for explicit start and stop commands when transmitting burst signals.

A particular challenge with streaming \ac{Tx} is handling errors:
In the \ac{CHDR} protocol used in the \acp{USRP}, this is done by monitoring the sequence numbers of incoming packets~\cite{RFNoCSpec}.
Errors are reported to the host and \ac{Tx} operation is only resumed after explicit acknowledgement.
The start and end of burst must be explicitly marked to allow discrimination between an intentional interruption and packet loss.
Correctly handling the loss of these delimiter packets is particularly complex~\cite{engelhardt2022lowlatency}.
In our application, due to limited resources and high data rates, the buffers on the \ac{FPGA} cannot be realized large enough to have sufficient slack to allow for the retransmission of lost packets.

We have therefore opted for a different approach, significantly simplifying the protocol design:
The header of each sample packet indicates its desired playback time expressed as beatcounter value.
The \ac{FPGA} offers multiple statistics counters, capturing the number of successfully transmitted packets as well as the number of packets discarded due to late arrival.
The host regularly reads these counters via \ac{AXIoE}, whereby the \ac{FPGA} attaches the timestamp when exactly the counter values were last changed.
Since the software knows how many sample packets should have been consumed by the \ac{DAC} at any given time, it can determine whether packets have been lost.

This solution also keeps the gap in the \ac{Tx} signal as small as possible:
When a packet is lost, only the samples in the affected packet are missing and the output auto-zero on idle feature implicitly handles the error on the \ac{FPGA} side.
Consecutive packets are played as intended by their beatcounter target.
In addition, there is no delay caused by waiting for explicit acknowledgments.

\subsection{\glsfmtshort{FPGA}-Design Realization}

\subsubsection{Loading-Looping-Approach}

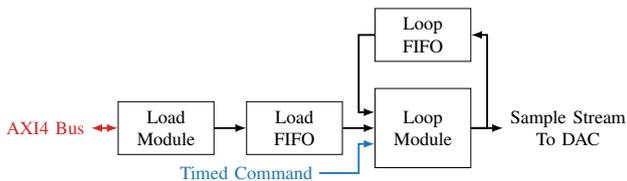
\begin{figure}[bt!]
  \centering
  \scalebox{0.7}{\begin{tikzpicture}

    \definecolor{color0}{rgb}{0.12156862745098,0.466666666666667,0.705882352941177}
    \definecolor{color1}{rgb}{1,0.498039215686275,0.0549019607843137}
    \definecolor{color2}{rgb}{0.172549019607843,0.627450980392157,0.172549019607843}
    \definecolor{color3}{rgb}{0.83921568627451,0.152941176470588,0.156862745098039}
    \definecolor{color4}{rgb}{0.580392156862745,0.403921568627451,0.741176470588235}
    \definecolor{color5}{rgb}{0.549019607843137,0.337254901960784,0.294117647058824}

    \tikzset{
       boxa/.style={rectangle, draw =black,minimum width = 1.8cm, minimum height = 1cm,align = center}
    }

    \node [boxa](LoadM){Load\\Module};
    \node [right = .6cm of LoadM] [boxa](LoadF){Load\\FIFO};				
    \node [right = .6cm of LoadF] [boxa, minimum height = 1.5cm](LoopM){Loop\\Module};
    \node [above = .5cm of LoopM] [boxa](LoopF){Loop\\FIFO};

    \node [below = .1cm of LoadF.south west][color0] (tc){Timed Command};
    \node [left = .5cm of LoadM][color3] (axi){AXI4 Bus};
    
    \node at ($(LoopM.east) + (.3,0)$)(align0) {};
    \node at ($(LoopF.east) + (.3,0)$)(align1) {};
    \node at ($(LoopM.west) + (-.3,.3)$)(align2) {};
    \node at ($(LoopF.west) + (-.3,0)$)(align3) {};
    \node at ($(LoopM.west) + (-.3,-.3)$)(align4) {};

     \draw[color0,->,line width = 1pt](tc.east) -| (align4.center) -- (LoopM.west|-align4.center);
     \draw[color3,<->,line width = 1pt](axi) -- (LoadM);

    \draw[->,line width = 1pt](LoadM) -- (LoadF);
    \draw[->,line width = 1pt](LoadF) -- (LoopM);
    \draw[->,line width = 1pt](LoopM.east) -- ++ (0.6,0) node [right, align =center]{Sample Stream\\To DAC};
    \draw[->,line width = 1pt](LoopM) -- (align0.center) -- (align1.center) -- (LoopF);
    \draw[->,line width = 1pt](LoopF) -- (align3.center) -- (align2.center) -- (LoopM.west|-align2.center);

\end{tikzpicture}}
  \caption{
    \ac{Tx} loading-looping \ac{FPGA} path: Conversion of \ac{AXI} writes to \ac{AXI}-Stream packets. Timed command-based loading-looping of sequences. Lightweight approach allows for playback of periodic signals, eliminating the bottleneck of the link to the host.}
  \label{fig:FPGATxLooping}
\end{figure}

\autoref{fig:FPGATxLooping} visualizes the architecture of the loading-looping \ac{DAC} data path:
Using \ac{AXIoE}, the host writes the desired sequence into the loader module, which passes the sequence as an \ac{AXI}-Stream packet to the load \ac{FIFO}.
After loading is finished, the host may control the playback of the sequence using timed commands.
The playback module outputs the sequence as sample stream to the \ac{DAC}, but also feeds it back to its input via the loop \ac{FIFO} for repetitive playback.

\subsubsection{Real-time \glsfmtshort{Tx} Streaming}

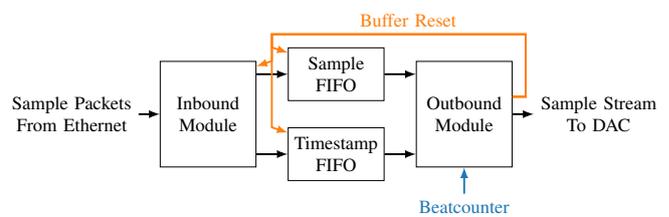
\begin{figure}[bt!]
  \centering
  \scalebox{0.7}{\begin{tikzpicture}

    \definecolor{color0}{rgb}{0.12156862745098,0.466666666666667,0.705882352941177}
    \definecolor{color1}{rgb}{1,0.498039215686275,0.0549019607843137}
    \definecolor{color2}{rgb}{0.172549019607843,0.627450980392157,0.172549019607843}
    \definecolor{color3}{rgb}{0.83921568627451,0.152941176470588,0.156862745098039}
    \definecolor{color4}{rgb}{0.580392156862745,0.403921568627451,0.741176470588235}
    \definecolor{color5}{rgb}{0.549019607843137,0.337254901960784,0.294117647058824}

    \tikzset{
       boxa/.style={rectangle, draw =black,minimum width = 1.8cm, minimum height = 1cm,align = center}
    }

    \node [boxa, minimum height = 2cm](IB){Inbound\\Module};
    \node [right = 3cm of IB] [boxa, minimum height = 2cm](OB){Outbound\\Module};
    \node [below = 0.25cm of $(IB)!0.5!(OB)$][boxa](TrigBuff){Timestamp\\FIFO};
    \node [above = 0.25cm of $(IB)!0.5!(OB)$][boxa](JittBuff){Sample\\FIFO};
    
    \draw[<-,line width = 1pt](IB.west) -- ++(-0.4,0)node [left, align=center]{Sample Packets\\From Ethernet};
    \draw[->,line width = 1pt](IB.east|-JittBuff.west) -- (JittBuff.west);
    \draw[->,line width = 1pt](IB.east|-TrigBuff.west) -- (TrigBuff.west);
    \draw[<-,line width = 1pt](OB.west|-JittBuff.east) -- (JittBuff.east);
    \draw[<-,line width = 1pt](OB.west|-TrigBuff.east) -- (TrigBuff.east);
    \draw[->,line width = 1pt](OB.east) -- ++ (0.4,0) node [right, align =center]{Sample Stream\\To DAC};

    \node at ($(JittBuff.north west)+(-0.3,0)$)(align0){};
    \node at ($(TrigBuff.north west)+(-0.3,0)$)(align1){};

    \draw[color0,<-,line width = 1pt](OB.south) -- ++ (0,-0.5) node[below]{Beatcounter};
    
    \draw[color1,->,line width = 1pt](OB.20) -- ++ (0.25,0) |- ($(JittBuff.north)+(0,0.25)$) node[above,pos=0.81] {Buffer Reset} -| (align1.center) -- ($(TrigBuff.north west)+(0,-0.1)$);
    \draw[color1,->,line width = 1pt](OB.20) -- ++ (0.25,0) |- ($(JittBuff.north)+(0,0.25)$) -| (align0.center) -- ($(JittBuff.north west)+(0,-0.1)$);
    \draw[color1,->,line width = 1pt](OB.20) -- ++ (0.25,0) |- ($(JittBuff.north)+(0,0.25)$) -| (align0.center |- IB.north) -- ($(IB.north east)+(0,-0.1)$);

\end{tikzpicture}}
  \caption{
    \ac{Tx} streaming \ac{FPGA} path: Beatcounter-based payback of inbound sample packets based on \ac{AXI}-Stream. Includes protocol-based error detection and handling by buffer reset. Fire-and-forget protocol design reduces complexity and resource consumption of the \ac{FPGA} realization.}
  \label{fig:FPGATxStreaming}
\end{figure}

The Tx streaming path design is shown in \autoref{fig:FPGATxStreaming}.
The inbound module performs a sequence check and inserts samples and timestamps into individually buffered sub-paths.
The outbound module manages time-controlled playback and outputs zero-sample words when no sample data are available.
Inbound and outbound module provide timestamped statistics counters, which can be read out by the host via \ac{AXIoE}.

The architecture implements a global packet status check, which ensures packet bit integrity.
On protocol level, four types of errors may occur and are handled appropriately:
\begin{enumerate}
  \item Packet loss is implicitly handled by the auto-zero on idle feature.
  \item An out-of-order packet is handled by the inbound module by discarding the late packet.
  \item A sample \ac{FIFO} underrun occurs when packets arrive too late.
  The outbound module recognizes a missed target time and triggers a reset of both \acp{FIFO}.
  The inbound module monitors the reset condition.
  \item A sample \ac{FIFO} overflow occurs when too many packets arrive too early.
  The inbound module detects and safely resolves backpressure conditions at both \ac{FIFO} inputs.
  This ensures sub-path synchronization.
\end{enumerate}
After any error condition occurred and has been resolved, the inbound module discards inbound data until it resyncs its input interface to the next Ethernet packet.
This design guarantees the shortest possible interruption of the output sample stream.

\subsection{Host-side Implementation}
The loading-looping approach places no special demands on the host -- the sequence to be played back is loaded into the device by a driver call from the user application (e.g., a Python script) using the \ac{AXIoE} control path.

\ac{Tx} streaming is more complex:
Here, the host driver ensures that the relatively small buffers in the \ac{FPGA} neither overflow nor underflow.
As described in \autoref{pacing}, we use the \gls{NIC}'s send on timestamp feature for this purpose.
To detect errors, the host reads the previously described status counters periodically via \ac{AXIoE}.
Based on the timestamp attached to the result, the host driver calculates how many sample packets are expected to have been consumed by the DAC.
From this, the host determines how many errors of which type have occurred during the readout interval.

\section{Implementation Example}\label{sec:impl}
The proposed \ac{SDR} architecture, described in the previous sections, is not tied to any particular hardware.
Starting from this section, we introduce a specific implementation (based on the Xilinx RFSoC).
This allows us to discuss relevant implementation details and validate our architecture with real measurements.

\subsection{Xilinx Zynq Ulrascale+ RFSoC XCZU48DR}
Besides the actual \ac{FPGA}, the Xilinx RFSoC XCZU48DR monolithically integrates specialized blocks, e.g., \SI{100}{\giga\bit\per\second} Ethernet, ADC, and DAC:

\begin{table}[h]
	\centering
	\caption{Xilinx XCZU48DR: \glsfmtshort{ADC} characteristics.~\cite{ds889,pg269}}
	\begin{tabular}{p{.5\linewidth}|p{.2\linewidth}}
		\toprule
    \textbf{Number of channels} & 8\\
		\textbf{Interleaved sub-ADCs per channel}& 8\\
    \textbf{Resolution} & 14\,bit\\
		\textbf{Sample rate} & 1--5\,GSa/s\\
  	\textbf{Analog bandwidth (-3dB)} & 6\,GHz\\
    \bottomrule
	\end{tabular}
	\label{tab:ADCSpecs}
\end{table}

\begin{table}[h]
	\centering
	\caption{Xilinx XCZU48DR: \glsfmtshort{DAC} characteristics.~\cite{ds889}}
  \begin{tabular}{p{.5\linewidth}|p{.2\linewidth}}
		\toprule
		\textbf{Number of channels}& 8\\
		\textbf{Resolution} & 14\,bit\\
		\textbf{Sample rate} & 0.5--9.85\,GSa/s\\
		\textbf{Analog bandwidth (-3dB)} & 6\,GHz\\
    \bottomrule
	\end{tabular}
	\label{tab:DACSpecs}
\end{table}

The \ac{RFdc} hard-IP offers direct RF-sampling \acp{ADC} and \acp{DAC}.
\autoref{tab:ADCSpecs} and \autoref{tab:DACSpecs} list their parameters.
Both converter types integrate digital signal processing features like \ac{DDC} and \ac{DUC}.
The \acp{ADC} and \acp{DAC} are organized in tiles.
Their power up sequence is neither synchronized nor timed, which results in an undeterministic timing relationship between tiles of a single as well as multiple devices.
The converter \ac{MTS} procedure ensures a consistent and deterministic timing across all \ac{ADC} and \ac{DAC} tiles.
This requires specific external clocks and sync signals, which have to be reconfigured multiple times.
All converter features are accessible through a single IP core, which can be customized for the design and provides data as well as control ports.

Xilinx provides a \SI{100}{\giga\bit\per\second} Ethernet interface by combining a CMAC hard IP with a GTY serial transceiver quad.
The RFSoC features two\footnote{Currently, we are limited to a single CMAC due to evaluation board layout constraints.} of these interfaces.

Beside the \ac{FPGA} as \ac{PL}, the RFSoC integrates a \ac{CPU} core as \ac{PS}, which in our case runs a Linux system and initializes the platform by configuring peripheral clock generation components\footnote{The clock generation network's communication interface is hardwired to the \ac{PS} package pins on the used evaluation board.} and loading the bitstream onto the \ac{FPGA}.
\ac{AXI} interfaces between \ac{PL} and \ac{PS} allow communication between both sides.
By connecting \ac{AXIoE} to the \ac{PS}-\ac{PL} interface and utilizing a driver in the \ac{PS}, the host is empowered to remotely reconfigure all peripheral components that are accessible via the \ac{PS}.

\subsection{Host}
On the host side, enterprise \ac{COTS} hardware is used.
Particularly noteworthy is the SSD array, which allows continuous recording of one channel's data at its full sample rate of \SI{5}{\giga\sample\per\second}:
It consists of 4x \textit{Samsung SSD PM9A3} with \SI{4}{\giga\byte\per\second} write rate each, so that a RAID0 reaches a sustained write rate of \SI{16}{\giga\byte\per\second}.
The \SI{100}{\giga\bit\per\second} Ethernet \gls{NIC} used to communicate with the \ac{FPGA} is an \textit{Nvidia Mellanox MCX623106AN-CDAT}.
It offers the required feature send-on-timestamp (also called packet pacing), allowing for the exact timing of packet transmission~\cite{DPDK_NIC}.
Both the SSD array and the \gls{NIC} are connected via PCI Express 4.0 directly to the CPU for maximum transfer rate.

\subsection{Device Clocking}
As \ac{REFCLK}, we use a \SI{100}{\mega\hertz} square wave instead of the common \SI{10}{\mega\hertz} sine wave to achieve a better phase noise performance.
A 1\,PPS reference signal is fed to the converter device.
Based on these, each device derives its clocks and \ac{SYSREF}.
This enables synchronous sampling and absolute synchronization in multi-device setups.

\subsection{RFdc-over-Ethernet}

\begin{figure}[bt!]
  \centering
  \scalebox{0.7}{\begin{tikzpicture}

    \definecolor{color0}{rgb}{0.12156862745098,0.466666666666667,0.705882352941177}
\definecolor{color1}{rgb}{1,0.498039215686275,0.0549019607843137}
\definecolor{color2}{rgb}{0.172549019607843,0.627450980392157,0.172549019607843}
\definecolor{color3}{rgb}{0.83921568627451,0.152941176470588,0.156862745098039}
\definecolor{color4}{rgb}{0.580392156862745,0.403921568627451,0.741176470588235}
\definecolor{color5}{rgb}{0.549019607843137,0.337254901960784,0.294117647058824}

    \tikzset{
       boxa/.style={rectangle,draw=black, top color=white, inner sep=5pt,minimum width=1.2cm, minimum height=0.4cm, text centered, text width=2.5cm},
       arw/.style={>={Triangle[length=5mm,width=8mm]},line width=5mm,draw=lightgray},
       axi/.style={draw=color3,line width=1pt}
    }

        \node[boxa] (axihost) {AXIoE Client};

        \node[fit=(axihost), draw, black, rectangle,label={[name=dpdklabel]above:DPDK-Based driver}] (dpdk) {};

    \node[boxa, left=of axihost] (py)  {Python Application};
    \node[boxa, above=of py] (rfdcdriver)  {RFdc Driver};

    \draw[<->] (py) -- (axihost);
    \draw[axi, <->] (rfdcdriver) -- (py);

    \node[fit=(rfdcdriver)(dpdk), inner sep=0.25cm, draw, black, rectangle,label=above:Host] (host) {};

    \node[boxa, below=2cm of axihost] (axifpga) {AXIoE Server};

    \node[boxa, left=of axifpga] (rfdc) {RFdc Configuration Interface};

    \node[boxa, below=of rfdc, opacity=0.4, rectangle,draw] (rfdcps) {RFdc Driver};
    \node[boxa, below=of rfdc, opacity=0.4, rectangle, cross out,draw] (rfdcps) {RFdc Driver};

    \node[fit=(axifpga)(rfdcps)(rfdc), inner sep=0.25cm, draw, black, rectangle,label=below:RFSoC device] (rfsoc) {};

    \coordinate (sep) at ($(rfdcps.north)!0.5!(rfdc.south)$);
    \draw[] (rfsoc.west|-sep) -- (rfsoc.east|-sep);

    \coordinate (a) at ($(rfsoc.west|-sep)!0.75!(rfsoc.east|-sep)$);
    \node[above=0.1cm of a] (pl) {PL (FPGA)};
    \node[below=0.1cm of a] (ps) {PS (ARM Core)};

    \draw[axi, <->, opacity=0.4] (rfdc) -- (rfdcps);
    \draw[axi, <->] (rfdc) -- (axifpga);
    \draw[axi, <->] (axihost) -- node[pos=0.4, left] (eth) {Ethernet} (axifpga);
    \draw[axi, <->] (axihost) -- (py);




\end{tikzpicture}}
  \caption{The configuration interface of the RF data converters is typically operated out of the PS, an integrated ARM core. Instead, we connect it to the host via our AXI over Ethernet infrastructure. There, the driver is linked to the Python application, allowing it to configure the hardware with maximum flexibility.}
  \label{fig:rfdcoe}
\end{figure}
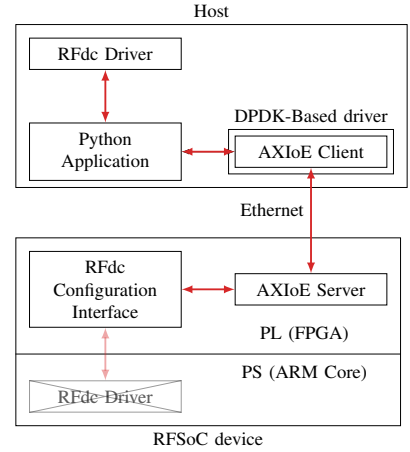

In addition to the sample transfer interface, the \ac{RFdc} IP core in the \ac{FPGA} design provides an \ac{AXI} control interface.
It is used, e.g., for initialization, calibration, and synchronization.
To access this interface, Xilinx provides a driver which is intended to run on the \ac{PS}.
This contradicts the goal of our \ac{SDR} architecture to have the host control the device as comprehensively as possible.

Therefore, we decided to give the host direct control over the \ac{RFdc} by attaching its configuration interface to the \ac{AXIoE} \ac{AXI} network as depicted in \autoref{fig:rfdcoe}.
However, the control registers of the \ac{RFdc} are not documented publicly, so the original driver must be utilized to interact with them.
Fortunately, it is published in C source code, so it is possible to compile it for the host architecture.
We adapted the functions that access the memory-mapped control register in the PS to trigger \ac{AXIoE} transactions instead.
This is transparent to the driver, as the underlying \ac{AXIoE} layer handles any problems that may occur, such as packet loss.

The only remaining hurdle is that the driver is designed for synchronous register accesses, as from the PS this is a fast local access, but the Python application follows the paradigm of asynchronous programming.
The solution here is provided by the library \textit{ucontext}, which makes it possible to create a separate execution context (especially stack) for the driver.
Its execution is suspended for every \ac{AXIoE} access and continued after the asynchronous arrival of the result.

The concept of tunneling accesses and commands through the reliable and ordered \ac{AXIoE} interface is not limited to the \ac{RFdc} and its driver.
Instead, it is generically applicable for most commercial IPs, custom modules, and even hardware peripherals, which are controlled by software drivers via register accesses.
Beyond that, \ac{AXIoE} can easily be utilized as reliable tunnel for non-\ac{AXI} interfaces via wrapper modules and on top benefits from the timed command functionality already included in the design.

\section{Measurement Results}\label{sec:txeval}
After implementing the proposed system architecture, we carried out a multitude of tests to verify its performance.

First, we examined the stability of the synchronization and command timing:
To do this, we power cycled the device numerous times and performed all synchronization steps, which include the \gls{MTS}, repeatedly.
In each cycle, we measured the system latency in an analog loopback.
Using timed commands, we set up the DAC path to produce a periodic test signal.
The generated analog Tx signal was looped back into the Rx channels, the ADC samples were recorded, and the delay between transmitted and received signal was determined.
Within each DAC/ADC combination, the measured latency was constant across multiple reboots.
This not only confirms that the synchronization, both within the FPGA design and between the data converters, works deterministically, but also validates proper command timing.

Next, we evaluated our control plane implementation in high-rate switching scenarios similar to, e.g., MIMO channel sounding~\cite[Fig. 2]{DanielUSRPSwitching}.
The software not only has to produce packets fast enough, but also pace them precisely so as not to overload the command queues in the converter device.
We verified that commands were reliably executed even when sending \ac{AXIoE} packets with the high pace of \SI{5}{\micro\second}.
This proves that even fast switching tasks do not have to be implemented explicitly in the FPGA but can be handled by the host in software.

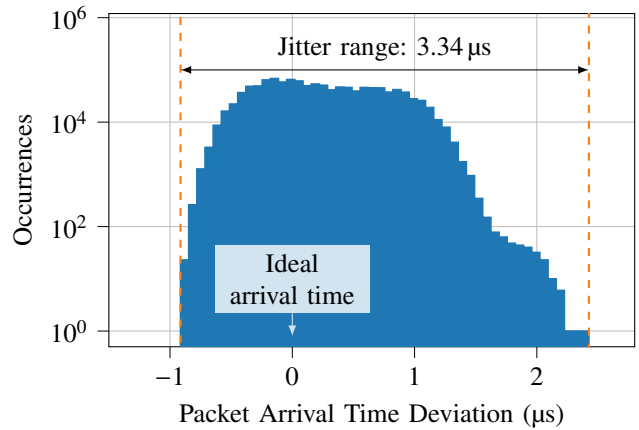
\begin{figure}[bt!]
\centering
\begin{tikzpicture}

\definecolor{color0}{rgb}{0.12156862745098,0.466666666666667,0.705882352941177}
\definecolor{color1}{rgb}{1,0.498039215686275,0.0549019607843137}

\begin{axis}[
log basis y={10},
tick align=outside,
tick pos=left,
xlabel={Packet Arrival Time Deviation (µs)},
xmajorgrids,
xmin=-1.5, xmax=2.8,
xtick style={color=black},
ylabel={Occurrences},
ymajorgrids,
ymin=0.5, ymax=1200000,
ymode=log,
ytick style={color=black},
]

\addplot [semithick, color0, const plot mark right,name path=curve]
table[x expr= -(\thisrow{slack}-5), y= occurrences] {%
slack occurrences
2.5742 0.00001
2.641 1
2.7078 1
2.7746 1
2.8414 6
2.9082 10
2.975 23
3.0418 32
3.1086 40
3.1754 44
3.2422 48
3.309 63
3.3758 78
3.4426 150
3.5094 346
3.5762 947
3.643 1743
3.7098 4095
3.7766 7983
3.8434 11128
3.9102 19053
3.977 25955
4.0438 28052
4.1106 37404
4.1774 41944
4.2442 37603
4.311 45209
4.3778 45469
4.4446 46133
4.5114 39389
4.5782 46146
4.645 46928
4.7118 41342
4.7786 50854
4.8454 53748
4.9122 49739
4.979 61368
5.0458 65495
5.1126 58566
5.1794 68438
5.2462 64359
5.313 49282
5.3798 48433
5.4466 36746
5.5134 22226
5.5802 16233
5.647 8708
5.7138 3286
5.7806 1270
5.8474 261
5.9142 23
5.981 0.00001
};

    \addplot[draw=none,name path=B] {0.0001};
    \addplot[color0] fill between[of=curve and B];

\draw[draw=color1, thick, dashed] (axis cs: 2.4258,0.1) rectangle (axis cs: -0.9142,10000000);
\draw[<->] (axis cs: 2.4258,100000) -- (axis cs: -0.9142,100000) node[pos=0.5,above] {Jitter range: \SI{3.34}{\micro\second}};


\node[text width=1.8cm, fill=color0!20, align=center] (desc) at (axis cs: 0,10) {Ideal \\arrival time};
\draw[color=color0!20, ->] (desc.south) -- (axis cs: 0,0.8);

\end{axis}

\end{tikzpicture}
\caption{
Histogram of packet arrival time deviation, positive values indicate late packet arrival. 
Using hardware send-on-timestamp, the time at which packets arrive at the device can be planned with a jitter of only \SI{3.34}{\micro\second}.
This accuracy allows implementing \ac{Tx} streaming with very small buffers in the \ac{FPGA}, conserving device resources.
}
\label{fig:TimingSlackHist}
\end{figure}

To verify the performance of the send-on-timestamp mechanism, we measured the arrival time jitter.
We did this by regularly inserting \ac{AXIoE} requests into the data stream, which command the \ac{FPGA} to read out the current value of the time stamp counter.
As shown in \autoref{fig:TimingSlackHist}, the range in which the real arrival time differs from the planned arrival time (jitter) spans \SI{3.34}{\micro\second}.
This proves the precision of both the hardware send-on-timestamp feature and clock modeling, confirming our design decision to keep \ac{Tx} streaming and control plane buffers on the device as small as possible.

Next, we evaluated the sustained performance of the Rx path:
Over several hours of continuous sample streaming at the full rate of \SI{5}{\giga\sample\per\second}, no packet loss occurred.
Also, continuously storing the incoming data stream of \SI{10}{\giga\byte\per\second} to the SSD array has been proven to work reliably, only limited by the storage capacity.

We successfully verified the timing and synchronization of both \ac{Tx} path variants.
\ac{Tx} streaming was demonstrated to work reliably at \SI{5}{\giga\sample\per\second}, i.e., a net data rate of \SI{10}{\giga\byte\per\second}.
Hereby, \SI{112}{\micro\second} of sample buffer were required on the device, which is more than we expected.
This is not a matter of the architecture itself, but most likely caused by a problem with the utilized host hardware, which is currently under investigation~\cite{DPDKUsers}.
Nevertheless, at a sample rate of \SI{2.5}{\giga\sample\per\second}, this effect disappeared and only \SI{14}{\micro\second} of samples had to be buffered on the device for reliable \gls{Tx} streaming.

Finally, to verify the real-time capabilities of our platform, we measured the round-trip latency in the following test scenario:
For both Rx and Tx, we set up a single continuous stream at a rate of \SI{2.5}{\giga\sample\per\second}.
To ensure that each Tx sample packet arrives in time, the software must enqueue it into the \ac{NIC} transmit queue with a fixed lead time.
We identified \SI{54}{\micro\second} as the minimum viable value for reliable operation.
Connecting two converters in an analog loop and analyzing the respective timestamps, we determined a latency of \SI{0.12}{\micro\second} from the DAC's digital input to the ADC's digital output.\footnote{This latency is primarily caused by the \ac{RFdc} IP core.}
In the Rx path, a sample reaches the software at most \SI{17.3}{\micro\second} after it was output by the ADC.
The individual values add up to the system's worst-case \textbf{round trip latency of 71.4\,µs}.

This makes our system ideally suited to meet the sub-millisecond latency requirements of state-of-the-art 6G applications~\cite{ericsson6g,huawei6g,nokia6g,samsung6g,zte6g}.
Therefore, it enables hardware-in-the-loop evaluation of each element of the communication system in a rapid prototyping environment:
From communication protocol components like scheduling, channel estimation, and beam steering to novel applications such as \gls{ISAC}.

\section{Application Example: 6G-ISAC-Radar}\label{sec:radar}
\Acl{ISAC} (\ac{ISAC}) is a proposed feature of the upcoming 6G standard~\cite{ITU2022,ITU2023,3gpp.6G.EU,3gpp.6G.US,Flagship6GSensing}.
Both infrastructure and sidelink-based communication may integrate radar-like sensing functionality~\cite{rosemannJCAS6G,WangICAS6G}.
This promises great benefits for use cases like assistance systems, health monitoring, mobility, public safety, and many more~\cite{3gpp.6G.EU,rosemannJCAS6G, Flagship6G-BW-UseCase}.

Owing to its unique features and performance (cf. \autoref{tab:ResultCompare}), our \ac{SDR} architecture is well suited for 6G rapid prototyping, e.g., for \ac{ISAC}~\cite{Flagship6GSensing}.
The standardization of \ac{ISAC} is at an early stage, i.e., comprising a work item~\cite{3gpp.5GNR.ISAC-WI} and case studies~\cite{3gpp.5GNR.ISACcasestudy,3gpp.5GNR.ISAC-channelmodel}.
Therefore, we opted for a straightforward application example:
A quasi-monostatic, single-input single-output Doppler radar demonstrator, inspired by the \ac{ISAC} features envisioned for 6G~\cite{rosemannJCAS6G}.
The example realizes arbitrary waveform generation and continuous sample acquisition and storage using our \ac{SDR} architecture.
The emitted signal uses the same orthogonal frequency-division multiplexing (OFDM) modulation scheme that is used by 5G and potentially 6G, representing illumination by a base station or user equipment~\cite{ITU2022}.
The architecture generally supports common mobile communication features, e.g., MIMO and beamforming, which can be implemented if appropriate multi-channel \ac{RF} frontends are available.
The real-time streaming capabilities with low latency and high bandwidth even facilitate the future implementation of advanced \ac{ISAC} demonstrators that seamlessly integrate into a 6G radio access network, serving as simultaneously sensing and communicating 6G network nodes.

\begin{figure}[bt!]
\centering
\scalebox{0.7}{\begin{tikzpicture}

    \tikzset{
bignode/.style={rectangle,draw=black, top color=white, inner    sep=1em,minimum width=1.75cm, minimum height=4.0cm, text centered},
mynode/.style={rectangle,draw=black, top color=white, inner sep=1em,minimum width=1.5cm, minimum height=1cm, text centered},
ifnode/.style={rectangle,dashed,draw=black, top color=white, inner sep=1em,minimum width=1.5cm, minimum height=1cm, text centered, fill=yellow},
fifonode/.style={rectangle,draw=black, top color=white, inner sep=1pt,minimum width=0.9cm, minimum height=0.4cm, text centered},
regnode/.style={rectangle,draw=black, top color=white, inner sep=5pt,minimum width=1.2cm, minimum height=0.4cm, text centered},
sarrow/.style={->, >=latex'},
carrow/.style={<->, >=latex', dashed},
branch/.style={fill,circle,minimum size=2pt,inner sep=0pt},
acaption/.style={below, midway, text width=1cm, align=center}
 }

    \node[regnode, label={[label distance=3mm, text width=2cm, text centered]above:5\,GSa/s \\per channel}] (tx1) {Tx};
    \node[mixer, scale=0.7] (txmixer) at ([xshift=2.5cm]tx1) {};
    \node[regnode, right=of txmixer] (txamp)  {Amplifier};
    \node[antenna, right=of txamp, yshift=0.5cm, scale=0.5] (txantenna) {};
    \node[waves, right=0.1cm of txantenna, yshift=0.5cm, rotate=-45] (txwaves) {};

    \draw[->] (tx1) -- (txmixer);
    \draw[->] (txmixer) -- (txamp.west);
    \draw[-] (txamp.east) -- (txantenna);

    \node[regnode, below=2cm of tx1] (rx1) {Rx};
    \node[mixer, scale=0.7] (rxmixer) at ([xshift=2.5cm]rx1) {};
    \node[regnode, right=of rxmixer] (rxamp)  {Amplifier};
    \node[antenna, right=of rxamp, yshift=0.5cm, scale=0.5] (rxantenna) {};
    \node[waves, right=1.5cm of rxantenna, yshift=1.5cm, xscale=-1, rotate=-45] (rxwaves) {};

    \node[text width=3cm, text centered] (DUT) at ([xshift=1.5cm]$(rxwaves)!0.5!(txwaves)$) {dynamic target\\\small{pedestrian, cyclist, ...}};

    \draw[<-] (rx1) -- (rxmixer);
    \draw[<-] (rxmixer) -- (rxamp.west);
    \draw[-] (rxamp.east) -- (rxantenna);

    \node[regnode] (logen) at ([xshift=-1cm, yshift=-2cm]rxmixer)  {LO Generator};
    \draw[->] (logen) |- ([yshift=-0.5cm]txmixer.south) -- (txmixer.south);
    \draw[->] (logen) |- ([yshift=-0.5cm]rxmixer.south) -- (rxmixer.south);

    \node (rfsoctext) at ([xshift=-1cm]$(tx1)!0.5!(rx1)$) {RFSoC};

    \node[fit=(tx1)(rx1)(rfsoctext), draw, black, rectangle] (rfsoc) {};

\end{tikzpicture}}
\caption{
Micro-Doppler radar measurement setup with a center frequency of $f_\text{c}=5$\,GHz, reaching an analog bandwidth of \SI{2}{\giga\hertz}, corresponding to a range resolution of \SI{7.5}{\centi\meter} in the monostatic case.}
\label{fig:RadarSetup}
\end{figure}
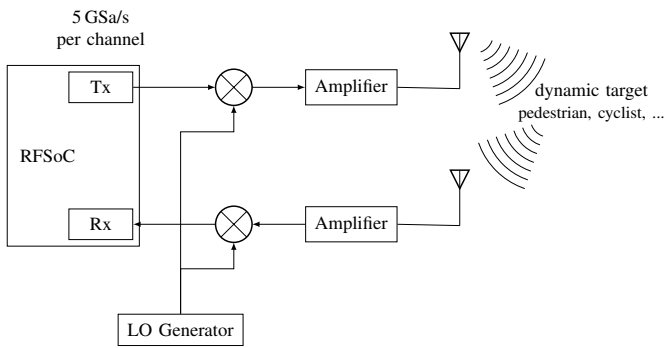

\begin{figure}[bt!]
  \centering
  \begin{subfigure}[t]{0.95\linewidth}
    \centering
    \resizebox{0.9\linewidth}{!}{\includegraphics{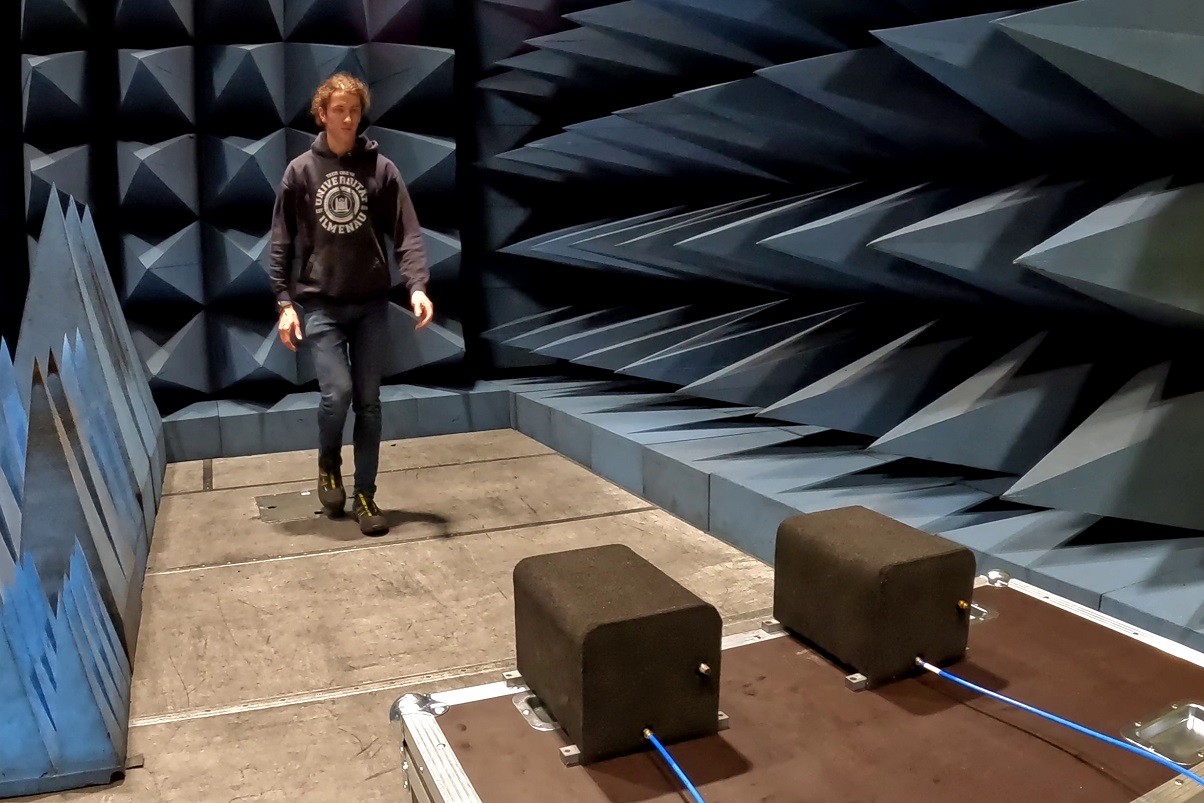}}
    \caption{
      Setup: Pedestrian walking towards our quasi-monostatic radar setup.}
    \label{subfig:MeasScenario}
  \end{subfigure}

  \begin{subfigure}[t]{0.95\linewidth}
    \centering
    \resizebox{\linewidth}{!}{
\begin{tikzpicture}

\definecolor{color0}{rgb}{0.12156862745098,0.466666666666667,0.705882352941177}
\definecolor{color1}{rgb}{1,0.498039215686275,0.0549019607843137}

\begin{axis}[
tick align=outside,
tick pos=left,
axis on top,
xlabel={Path length (m)},
xmajorgrids,
xtick style={color=black},
ylabel={Doppler speed (m/s)},
ymajorgrids,
xmin=6, xmax=15, ymin=-3, ymax=3,
ytick style={color=black},
colorbar,
colormap={bernd}{
    rgb=(0.96862745098039216,  0.98431372549019602,  1.0                )
    rgb=(0.87058823529411766,  0.92156862745098034,  0.96862745098039216)
    rgb=(0.77647058823529413,  0.85882352941176465,  0.93725490196078431)
    rgb=(0.61960784313725492,  0.792156862745098  ,  0.88235294117647056)
    rgb=(0.41960784313725491,  0.68235294117647061,  0.83921568627450982)
    rgb=(0.25882352941176473,  0.5725490196078431 ,  0.77647058823529413)
    rgb=(0.12941176470588237,  0.44313725490196076,  0.70980392156862748)
    rgb=(0.03137254901960784,  0.31764705882352939,  0.61176470588235299)
    rgb=(0.03137254901960784,  0.18823529411764706,  0.41960784313725491)
},
colorbar style={
    ylabel=Level (dBFS)},
point meta min=-25,
point meta max=-10,
]

    \addplot graphics [xmin=0, xmax=29.2968, ymin=-4.71859, ymax=4.71859, zmin=-25, zmax=-10] {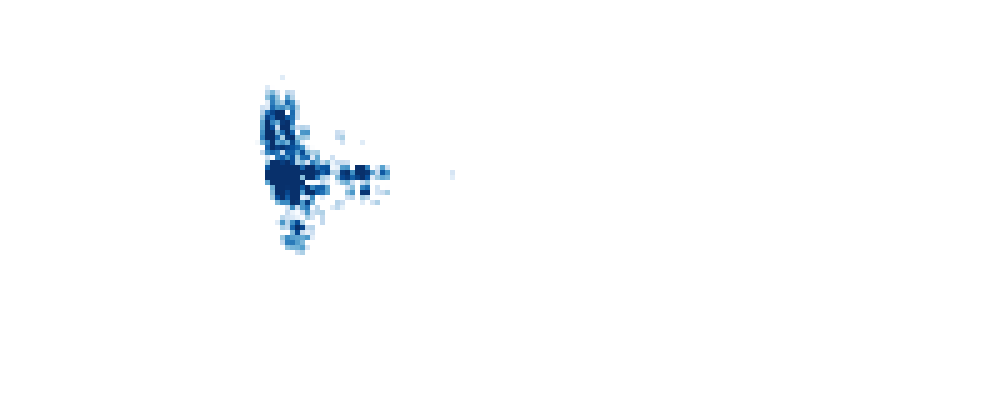};

\addplot [mesh] {sin(x)};
\end{axis}

\draw[<-, line width= 0.7pt] (2.4,3.6) --++ (.8,0) node[right,fill=white] {Arm + Leg} ;
\draw[<-, line width= 0.7pt] (2.6,1.7) --++ (.6,0) node[right] {Arm} ;
\draw[<-, line width= 0.7pt] (1.4,2.3) --++ (-.5,-.5) node[below] {Torso} ;
\end{tikzpicture}}
    \caption{
      Micro-Doppler signature: In the delay-Doppler plot, torso and extremities can be resolved. The proposed SDR system achieves a resolution (size of one pixel) of \SI{7.5}{\centi\meter} by \SI{11.8}{\centi\meter\per\second}.}
    \label{subfig:MeasPlot}
  \end{subfigure}
  \caption{
      \ac{ISAC} scenario: A pedestrian is captured by a radar sensor node implementing the proposed \ac{SDR} system architecture.}
    \label{fig:MeasExample}
\end{figure}
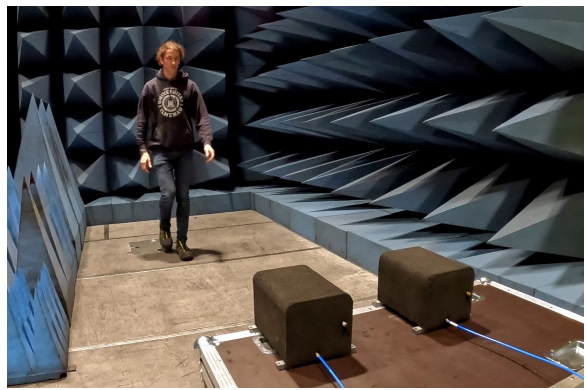
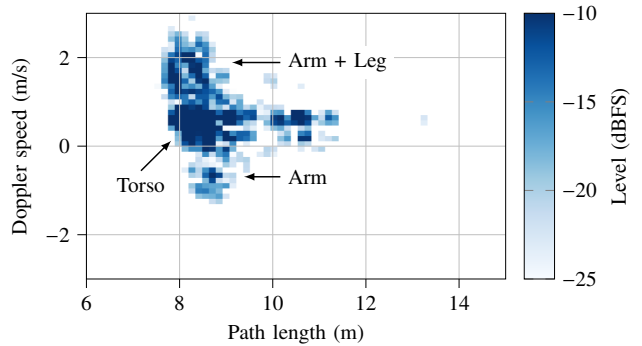

Regarding our simplified \ac{ISAC} application example, evaluating the Doppler effect over time and range will not only reveal the speed of a radar target as a whole but also the movements of its inner parts, the so-called micro-Doppler.
The entirety of a target's inner movements, e.g., a pedestrian's limbs or the individual rotor blades of a UAV, can be detected as characteristic patterns: Its micro-Doppler signature.
To resolve the target's inner structure in the range domain, a radar system with high instantaneous bandwidths is necessary.
With a bandwidth of \SI{2}{\giga\hertz}, our setup realizes a path distance resolution of \SI{15}{\centi\meter}, which corresponds to a range resolution of \SI{7.5}{\centi\meter} in the monostatic case.
In~\cite{RCSPaper}, we discuss the promising chances of joint evaluation of static and dynamic target reflectivity for detection, localization, and classification of targets in upcoming \ac{ISAC} solutions and therefore, future mobility applications.

\begin{table*}[t!]
  \centering
    \caption{Comparison of our solution with the state-of-the-art \ac{USRP} X440.~\cite{X440,X440kb}}
    \label{tab:ResultCompare}
      \setlength{\tabcolsep}{3pt}
      \begin{tabular}{r|p{6cm}|p{6cm}}
        \toprule
          &\textbf{UHD} & \textbf{This work} \\
          &\textbf{Ettus X440} & \textbf{XCZU48DR} \\
        \midrule
          Ecosystem & Mature & Recent\footnotemark[11]  \\ 
        \midrule
          \multicolumn{3}{l}{\textbf{Rx}  (cf. \autoref{sec:ADC})}\\[2pt]
          \makecell[tr]{Continuous, high-rate recording} & Must be implemented by user & Native support; demonstrated \SI{10}{\giga\byte\per\second} per \SI{100}{\giga\bit\per\second} link\\
          \makecell[tr]{Real-time sample access} & Must be implemented by user & Native support; decoupled, concurrent, zero-overhead access via shared memory\\
          Packet loss handling & Resynchronization must be implemented by user & Automatic zero-padding retains synchronization \\
          Sample distribution & A separate endpoint can be specified per stream & VLANs enable arbitrary distribution of samples to multiple hosts \\
        \midrule
        \multicolumn{3}{l}{\textbf{Tx} (cf. \autoref{sec:DAC})}\\[2pt]
          Arbitrary waveform generation & Flexible playback from SDR memory & Loading-looping via SDR memory\\
          Burst streaming & Separate handshake for each burst  & Native support enables zero overhead\\
          \makecell[tr]{Flow control mechanism \\(cf. \autoref{pacing})} & \makecell[tl]{Pause frames, blocking all traffic on interface\\(head-of-line blocking)} & Software-controlled per-stream packet pacing \\
          Packet loss handling & Resynchronization requires lengthy handshake~\cite{engelhardt2022lowlatency} & Auto-zero on idle inherently retains synchronization \\
        \midrule
        \multicolumn{3}{l}{\textbf{Hardware Control} (cf. \autoref{sec:Control})}\\[2pt]
          Synchronous parallel command execution & Limited due to shared command queues~\cite{DanielUSRPSwitching} & Full support via isolated command queues\\
          Fast automatic gain control (AGC) & Necessitates FPGA implementation~\cite{DanielUSRPSwitching} &  Demonstrated low latency enables implementation in software \\

        \midrule
        \multicolumn{3}{l}{\textbf{Performance Parameters}}\\[2pt]
          Maximum complex baseband sample rate & \SI{2}{\giga\sample\per\second}~\cite{X440kb} & \SI{2.5}{\giga\sample\per\second}\\
          Maximum instantaneous channel bandwidth & \SI{1600}{\mega\hertz}~\cite{X440kb} & \SI{2000}{\mega\hertz}\\
          Rx streaming rate (1x \SI{100}{\giga\bit\per\second} link) & \SI{4}{\giga\byte\per\second} (1 channel) -- \SI{6.4}{\giga\byte\per\second} (4 channels)~\cite{X440kb} & \SI{10}{\giga\byte\per\second}\\
          Tx streaming rate (1x \SI{100}{\giga\bit\per\second} link) & \SI{7.2}{\giga\byte\per\second}~\cite{X440kb} & \SI{10}{\giga\byte\per\second}\\
        \bottomrule
      \end{tabular}

\end{table*}

Our measurement setup is shown in \autoref{fig:RadarSetup}:
As only a single converter device is used, no multi-device synchronization is required.
It continuously transmits a complex-valued baseband OFDM sequence with a period length of 2500 samples and a bandwidth of \SI{2}{\giga\hertz}.
The baseband signal is mixed up into an \gls{RF} signal with a center frequency of \SI{5}{\giga\hertz}.
One antenna radiates the amplified signal while a second one receives the target's reflection.
This \gls{Rx} signal is amplified and mixed down into the baseband before being sampled by the converter device and sent to the host.
The resulting stream of \SI{10}{\giga\byte\per\second} is stored on the SSD array in real-time, while computing delay-Doppler plots in parallel.

As shown in \autoref{subfig:MeasScenario}, we adapted a typical \ac{ISAC} mobility scenario as measurement example:
A pedestrian is walking towards a concentrated radar sensor node, which is realized by our \ac{SDR} system and utilizes a quasi-monostatic antenna setup.
\autoref{subfig:MeasPlot} contains a snapshot of the continuously generated delay-Doppler plots.
The target's micro-Doppler signature can easily be identified as three characteristic peaks:
The pedestrian's torso appears with its walking speed as the peak in the center.
Moving relatively to the torso, the swinging arms appear centered around it as distinct peaks with different distances and speeds.
The signature's asymmetric shape, which is shifted towards higher absolute speeds, is caused by the stepping foot moving toward the antennas in the plotted time instant.
Due to the continuously changing target geometry, the plot's subsequent snapshots in time would show the arms' peaks moving around the torso's on elliptic curves.
This leads to the target's actual time-dependent micro-Doppler signature.

These measurements demonstrate a range resolution superior to that of state-of-the-art SDR solutions like the Ettus \ac{USRP} X440~\cite{X440kb}.
In the context of \ac{ISAC}-related research, we already used the system to measure micro-Doppler signatures of different kinds of targets relevant for future mobility scenarios \cite{RCSPaper} as well as to validate a target modeling algorithm for generating training data for artificial intelligence~\cite{BladeModelPaper}.

\footnotetext[11]{Interfaces to other scripting languages and GNU Radio can be easily developed.}
\FloatBarrier

\section{Future Work}\label{sec:FutureWork}

First, as shown in \autoref{sec:txeval}, a further latency improvement of the real-time \ac{Tx} streaming at \SI{5}{\giga\sample\per\second} is possible.
To accomplish this, a limitation associated with the specific host hardware must be overcome.
However, this is not a matter of the architecture itself and does not affect any application in which a static \ac{Tx} sequence is used as we successfully demonstrated.

Although already addressed in the architecture, one topic that we have not yet tested is the synchronization of multiple devices:
The first step is to synchronize multiple devices in a single node using a distributed clock in a wired setup.
Moving on to distributed multi-node measurement arrangements using GNSS as a time source is the next challenge.

We are also working on the integration of multiple \ac{RF} frontends, which will be used to access a variety of frequency bands.
For practical application in channel sounding, very high dynamic ranges are necessary, which is why we are working on an \acl{AGC}.

With applications demanding bandwidths exceeding the capability of a single converter channel, a transceiver has to extend its instantaneous bandwidth beyond the Nyquist limit of an individual channel.
To achieve this, channel bonding can be used.
In a test setup with reduced complexity based on the RFSoC, we already investigated and realized multiple approaches~\cite{sebpaper}.
They shall now be implemented and evaluated with our novel architecture.
Channel bonding requires multiple channels to be streamed from one or more converter devices to the host server in parallel.
This implies challenges concerning multi-channel and multi-device synchronization, the high-speed serial interface, and the host's real-time processing capabilities, which shall also be addressed in our future research.

Finally, a more precise evaluation of the RFSoC as a measurement system is also desirable:
The long-term stability of the \ac{RFdc}'s internal calibration needs to be investigated here.
Imperfections such as inter-channel imbalances should also be analyzed for possible correction via pre-distortion or post-processing.

\section{Conclusions}\label{sec:Conclusions}

This contribution proposes a novel generic and hardware independent \ac{SDR} system architecture, which covers functionality for \ac{Rx}, \ac{Tx}, and remote control.
It enables synchronized and distributed multi-node transceiver setups with high channel counts.
Both \ac{Rx} and \ac{Tx} implement time-triggered burst and continuous sample-streaming.
To enable periodic signal generation beyond the limits of host and data interface, the \ac{Tx} additionally features runtime loading-looping signal playback.
The control infrastructure realizes remote configuration, synchronization, and timed command execution.
In summary, the proposed system architecture covers the base functionality of an \ac{SDR} allowing for fast and easy high-level adaption of the realized platform to a variety of applications without requiring low-level changes to the \ac{FPGA} design or \Cpp code.
Therefore, we intrinsically support rapid prototyping.
Due to its modular, generic, and reconfigurable design, our architecture is ready for future extensions.

In order to validate the functionality of the design, we have realized our architecture on a Xilinx RFSoC XCZU48DR in combination with \ac{COTS} server hardware.
\Ac{Rx} streaming and recording as well as dynamic \ac{Tx} streaming were successfully implemented and demonstrated for continuous operation at the converters' maximum rate of \SI{5}{\giga\sample\per\second}.
For periodic Tx signals, the loading-looping design allows multi-channel high-rate operation exceeding the Ethernet interface data rate limit.

\autoref{tab:ResultCompare} compares our solution with the latest Ettus \ac{USRP} which is also based on Xilinx RFSoC technology.
The table shows that our solution achieves superior performance, leveraging on a better utilization of the available link data rate.
To the best of our knowledge, it is the only \ac{SDR} which allows to continuously send, receive, and record a signal with a sample rate of \SI{5}{\giga\sample\per\second} over a single \SI{100}{\giga\bit\per\second} Ethernet link.
At the same time, a latency of less than \SI{80}{\micro\second} is realized, which, to this date, no other software-based system achieves, and paves the way for sub-millisecond latency in 6G development.
Therefore, we bridge the gap between the high performance requirements of modern mobile communication and rapid application development.
Combining flexibility and performance, our SDR system architecture is ideal for research and applications requiring scalable and distributed RF transceiver solutions with high instantaneous bandwidths.
These include antenna measurements, radar target characterization, multi-node MIMO channel sounding, real-time algorithm testing, also incorporating AI-driven features, and ISAC in 6G.

\bibliographystyle{IEEEtran}
\bibliography{paper}

\begin{IEEEbiography}[{\includegraphics[trim={0cm 0.25cm 0.05cm 0.25cm},width=1in,height=1.25in,clip,keepaspectratio]{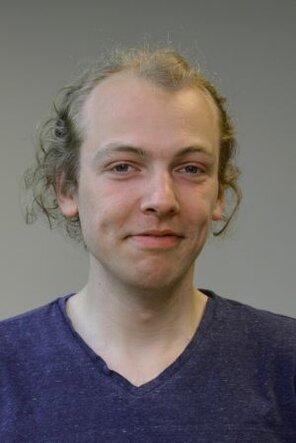}}]{M.Sc. Maximilian Engelhardt} received his Bachelor of Science (B.Sc.) and Master of Science (M.Sc.) degrees in computer and systems engineering from Technische Universität Ilmenau, Germany in 2019 and 2021.
Since 2021, he is working as research assistant at Fraunhofer IIS, Germany.
His research is centered around novel software-define radio solutions and radio-frequency measurement architectures, whereby his focus is on implementing real-time systems on off-the-self hardware.
\end{IEEEbiography}

\begin{IEEEbiography}[{\includegraphics[trim={.2cm 0.7cm .2cm .3cm},width=1in,height=1.25in,clip,keepaspectratio]{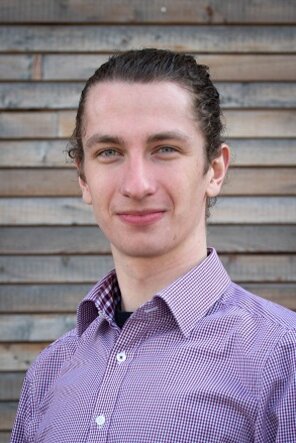}}]{M.Sc. Sebastian Giehl} has been working on parallel computing \ac{FPGA} systems since 2017.
He received his B.Sc. and M.Sc. degrees in electrical engineering and information technology from Technische Universität Ilmenau, Germany in 2020 and 2022.
In 2022, he has worked at Fraunhofer IIS, Germany.
Since 2023, he has been working as a research assistant for Technische Universität Ilmenau on \ac{ISAC} in mobility applications for the next generation of mobile communication and \ac{FPGA}-based, high-bandwidth, real-time capable \ac{SDR} systems for, e.g., radar or channel sounding.
\end{IEEEbiography}

\begin{IEEEbiographynophoto}{M.Sc. Michael Schubert} received his M.Sc. degree in Computer Engineering from Technische Universität Ilmenau in 2018.
Since then, he has been working on RFSoC-based measurement systems at the Fraunhofer Institute for Integrated Circuits IIS.
His research focuses on novel physical layer algorithms especially for the upcoming 6G mobile communication.
In particular, he works on FPGA implementations and related software components for signal processing.
\end{IEEEbiographynophoto}

\begin{IEEEbiography}[{\includegraphics[trim={0cm 0cm 0cm 0.0cm},width=1in,height=1.25in,clip,keepaspectratio]{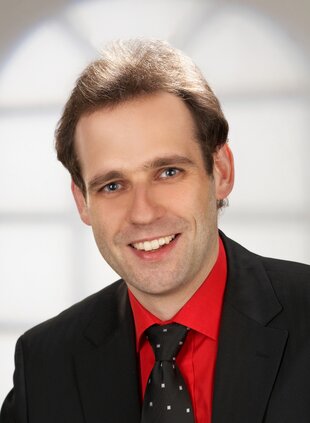}}]{Dr.-Ing Alexander Ihlow} received the Dipl.-Ing. degree in electrical engineering and information technology from Technische Universität Ilmenau, Germany, in 1999, and the Dr.-Ing. degree from Otto von Guericke University Magdeburg, Germany, in 2006.
Since 2008, he has been with the Institute of Information Technology, Technische Universität Ilmenau, as scientific staff member.
His current research interests include signal processing, wireless communications, and measurement and testing technology.
\end{IEEEbiography}

\begin{IEEEbiography}[{\includegraphics[trim={0.8cm 0.5cm 0.5cm 0.0cm},width=1in,height=1.25in,clip,keepaspectratio]{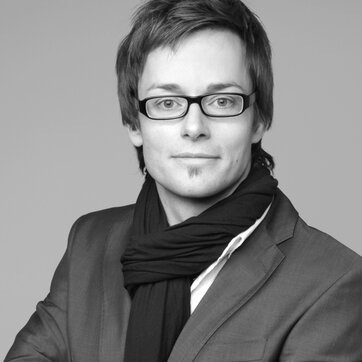}}]{Dipl.-Ing. Christian Schneider} received his Diploma degree in electrical engineering from the Technische Universität Ilmenau, Germany in 2001.
He is currently a group leader at the Electronic Measurements and Signal Processing department (EMS) at Technische Universität Ilmenau as well as at Fraunhofer IIS.
His research interests include multi-dimensional channel sounding, radio channel characterisation and modelling, and its application to space-time signal processing and \ac{ISAC} questions.
He received a best paper award at the European Wireless conference in 2013 and European Conference of Antennas and Propagation in 2017 and 2019.
\end{IEEEbiography}

\begin{IEEEbiography}[{\includegraphics[width=1in,height=1.25in,clip,keepaspectratio]{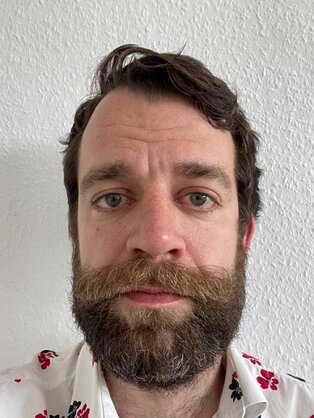}}]{M.Sc. Alexander Ebert} received the B.Sc. and M.Sc. degrees in electrical engineering and information technology from Technische Universität Ilmenau, Germany in 2012 and 2014, respectively.
Since 2021, he has been with the Electronic Measurements and Signal Processing Group at Fraunhofer IIS.
His current research interests include microwave and millimeter-wave systems, front-end design, and multiband and multichannel sounding up to THz frequencies.
\end{IEEEbiography}

\begin{IEEEbiography}[{\includegraphics[width=1in,height=1.25in,clip,keepaspectratio]{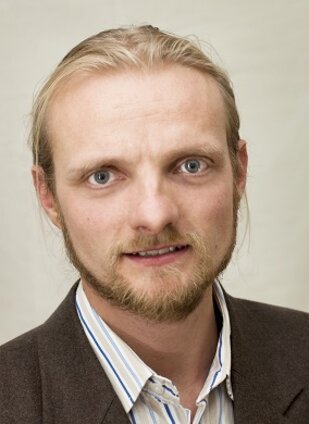}}]{Dr.-Ing. Markus Landmann}
received Dipl.-Ing. and Dr.-Ing. degrees in electrical engineering from Technische Universität Ilmenau, Germany in 2001 and 2008.
Until 2009, he worked as a research assistant at Technische Universität Ilmenau and Tokyo Institute of Technology on wireless propagation, channel modeling, and array signal processing.
Since 2010, he is working for Fraunhofer IIS responsible for the Facility for Over the Air Research and Testing including test and development of satellite and terrestrial based communication systems (2G-5G).
From 2018 he is Chief Scientist of the EMS Department being Co-department-head from 2022 and responsible for the strategic topics related to 5G standardization.
\end{IEEEbiography}

\begin{IEEEbiography}[{\includegraphics[width=1in,height=1.25in,clip,keepaspectratio]{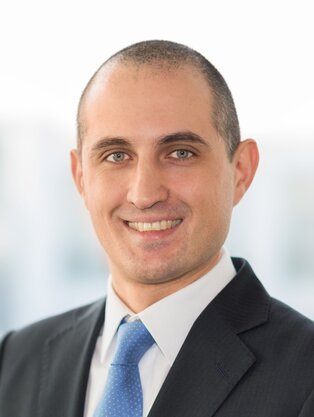}}]{Prof. Dr.-Ing. Giovanni Del Galdo} received the Laurea degree in telecommunications engineering from Politecnico di Milano, and the Dr.-Ing. degree on MIMO channel modeling from Technische Universität Ilmenau, Germany in 2007.
He then joined the Fraunhofer Institute for Integrated Circuits IIS, focusing on audio watermarking and spatial sound.
Since 2012, he leads a joint research group composed of a Department at Fraunhofer IIS and, as a Full Professor, a Chair with Technische Universität Ilmenau on the research area of electronic measurements and signal processing.
His current research interests include the analysis, modeling, and manipulation of multidimensional signals, over-the-air testing, and sparsity promoting reconstruction methods.
\end{IEEEbiography}

\begin{IEEEbiographynophoto}{M.Sc. Carsten Andrich} received the B.Sc. and M.Sc. degrees in electrical engineering and information technology from Technische Universität Ilmenau, Ilmenau, Germany, in 2014 and 2016, respectively.
Subsequently, he joined the Fraunhofer Institute for Integrated Circuits IIS as a researcher and engineered software-defined radio systems for integrated sensing and communication as well as radio propagation measurements.
Since 2020, he has been with the Institute of Information Technology, Technische Universität Ilmenau, as a scientific staff member.
There, he is the system architect responsible for the development of distributed, RFSoC-based measurement systems.
His area of expertise lies in distributed radio frequency measurements, software-defined radio systems, time and frequency synchronization, and digital signal processing.
\end{IEEEbiographynophoto}


\end{document}